\documentclass[lettersize,journal]{IEEEtran}
\usepackage{amsmath,amsfonts}
\usepackage{algorithmic}
\usepackage{algorithm}
\usepackage{array}
\usepackage[caption=false,font=normalsize,labelfont=sf,textfont=sf]{subfig}
\usepackage{textcomp}
\usepackage{stfloats}
\usepackage{url}
\usepackage{verbatim}
\usepackage{graphicx}
\usepackage{cite}

\usepackage[table]{xcolor} 
\usepackage{tabularx} 

\definecolor{greenCell}{rgb}{0.67, 0.88, 0.69}
\definecolor{yellowCell}{rgb}{0.8, 0.5, 0.1}

\begin{document}

\title{Physics-Informed LSTM-Based Delay Compensation Framework for Teleoperated UGVs}

\author{Ahmad~Abubakar, Yahya~Zweiri, 
AbdelGafoor~Haddad, Mubarak Yakubu,\\Ruqayya Alhammadi, and~Lakmal~Seneviratne

\thanks{Manuscript submitted February, 2024. This work was supported by Khalifa University under Award No. RC1-2018-KUCARS \it{(Corresponding author: Ahmad Abubakar.)}}
\thanks{A. Abubakar, A. Haddad, M. Yakubu, and R. Alhammadi are with the Department of Mechanical Engineering, Khalifa University of Science and Technology, Abu Dhabi, United Arab Emirates (e-mail: 100059792@ku.ac.ae, 100049699@ku.ac.ae, 100041348@ku.ac.ae).}
\thanks{Y. Zweiri and L. Seneviratne are with the Advanced Research and Innovation Center (ARIC) and Khalifa University Center for Autonomous Robotic Systems (KUCARS). They are also with the Department of Mechanical Engineering and the Department of Aerospace Engineering, respectively, Khalifa University of Science and Technology, Abu Dhabi, United Arab Emirates (e-mail: lakmal.seneviratne@ku.ac.ae, yahya.zweiri@ku.ac.ae).}
}

\markboth{Submitted to IEEE Transactions on Systems, Man, and Cybernetics, February~2024}%
{Shell \MakeLowercase{\textit{et al.}}: A Sample Article Using IEEEtran.cls for IEEE Journals}


\maketitle

\begin{abstract}

Bilateral teleoperation of low-speed Unmanned Ground Vehicles (UGVs) on soft terrains is crucial for applications like lunar exploration, offering effective control of terrain-induced longitudinal slippage. However, latency arising from transmission delays over a network presents a challenge in maintaining high-fidelity closed-loop integration, potentially hindering UGV controls and leading to poor command-tracking performance. To address this challenge, this paper proposes a novel predictor framework that employs a Physics-informed Long Short-Term Memory (PiLSTM) network for designing bilateral teleoperator controls that effectively compensate for large delays. Contrasting with conventional model-free predictor frameworks, which are limited by their linear nature in capturing nonlinear and temporal dynamic behaviors, our approach integrates the LSTM structure with physical constraints for enhanced performance and better generalization across varied scenarios. Specifically, four distinct predictors were employed in the framework: two compensate for forward delays, while the other two compensate for backward delays. Due to their effectiveness in learning from temporal data, the proposed PiLSTM framework demonstrates a 26.1\% improvement in delay compensation over the conventional model-free predictors for large delays in open-loop case studies. Subsequently, experiments were conducted to validate the efficacy of the framework in close-loop scenarios, particularly to compensate for the real-network delays experienced by teleoperated UGVs coupled with longitudinal slippage. The results confirm the proposed framework is effective in restoring the fidelity of the closed-loop integration. This improvement is showcased through improved performance and transparency, which leads to excellent command-tracking performance.
\end{abstract}


\section{Introduction}
\IEEEPARstart{T}{he} low-speed teleoperated Unmanned Ground Vehicles (UGVs) have gained significant attention in various fields, including industrial automation, military operations, and space exploration~\cite{b1,b2,b3}. These UGVs are typically used in scenarios that require careful navigation and interaction with the remote environment~\cite{b4}. A good example is the navigation of space rovers on soft terrains in planetary exploration~\cite{b5a}. 

Terrain traversability is one of the key challenges for teleoperated UGVs. In particular, soft terrains with some degree of roughness, such as regolith~\cite{b5aa}, pose unique challenges for UGV navigation due to the longitudinal slippage induced by wheel-terrain interaction. To enhance the navigation and ensure precise UGV control, the operator should be aware of the induced slippage and apply appropriate commands through a bilateral teleoperation system~\cite{b5b,b5c}.  Telepresence, which encompasses the operator's situational awareness, enables proactive and timely modification of control inputs, to optimize the terrain traversability and the UGV's stability and excellent command-tracking performance. Thus, haptic feedback could be a good approach to render slippage awareness in teleoperated UGV as it is difficult to detect slippage through video feedback alone~\cite{b6}. 

\begin{figure}[htbp]
\centering
\includegraphics[width=\columnwidth]{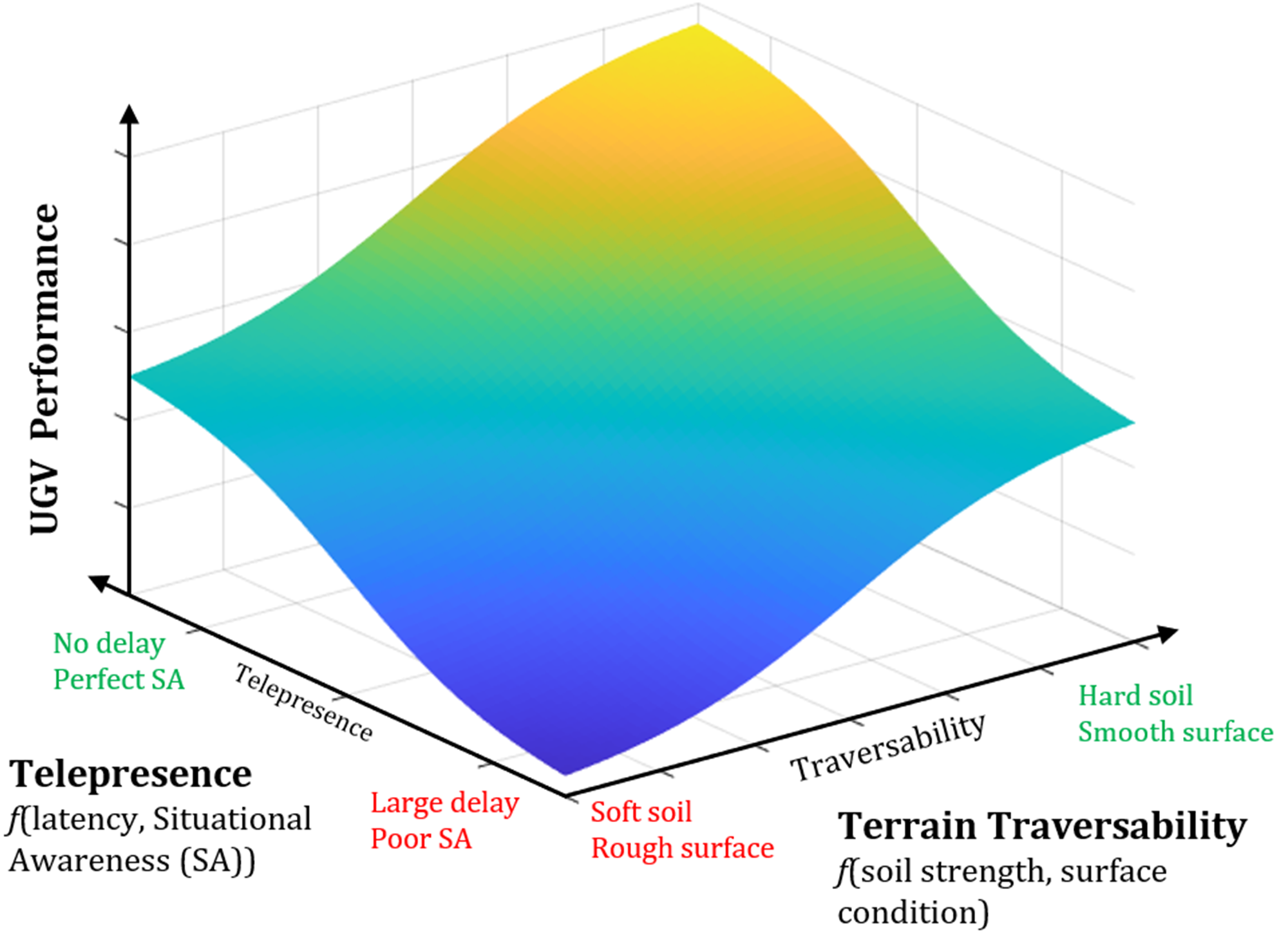}
\caption{The conceptual relationship illustrates how the command-tracking performance of teleoperated UGV is affected by both the terrain properties and the latency.}
\label{fig1}
\vspace{-15pt}
\end{figure}

However, latency (network delays) within the communication channel between the operator and the UGV can significantly affect telepresence by compromising the fidelity of the closed-loop integration necessary for timely responses to environmental changes~\cite{b3}. This delay may lead to degraded performance and transparency, with the potential to destabilize the system, particularly as the communication channel becomes nonpassive, even with small time delays~\cite{b7a,b7}. Meanwhile, the threshold for small delays that cause instability varies, depending on the specific system dynamics and application of the system~\cite{b8}. To address the challenges posed by latency in low-speed teleoperated UGVs subjected to induced longitudinal slippage, researchers have focused on various approaches including passivity-based, energy-based, and control-based strategies~\cite{b3}. Fig.~\ref{fig1} illustrates how varying degrees of telepresence and terrain traversability correlate with the overall performance of the teleoperated UGV.

In this paper, we examine control-based delay compensation techniques for low-speed teleoperated UGVs subject to longitudinal slippage in the presence of significant delays.

\subsection{Related Work}

Luz et al.~\cite{b9} highlight the importance of slippage awareness via haptic feedback in teleoperated UGVs operating in low-light conditions, showing improved control over induced slippage and faster task completion compared to relying on virtual feedback alone. Nonetheless, such discrete haptic feedback does not precisely convey precise slippage awareness to the operator~\cite{b4}. In~\cite{b10}, it is demonstrated that continuous haptic feedback, particularly based on the induced velocity loss of the UGV, can render slippage awareness more effectively than discrete feedback. Alternatively, the torque at the wheel contact point has been adopted to provide continuous haptic feedback for slippage awareness~\cite{b10b}, though it is not as effective as the velocity loss-based feedback. Consequently, these feedbacks significantly improve the control of induced longitudinal slippage. Building on~\cite{b10}, the authors further extended their approach in~\cite{b5b} by incorporating control of both lateral slide and longitudinal slippage simultaneously. However, a trade-off between command-tracking performance and system transparency was observed; specifically, enhancing performance sometimes introduced non-passivity issues due to negative slip conditions. To address these issues, a slippage compensator based on a feedforward controller was proposed~\cite{b5b}. Moreover, Li et al.~\cite{b12} addressed the non-passivity issue by modifying the slide angle in the UGV's kinematics and employing the Time Domain Passivity Approach (TDPA). Additionally, they proposed a slippage-dependent local controller in~\cite{b6}. Their strategies resulted in more effective control of both longitudinal slippage and lateral sliding, significantly enhancing the UGV's command-tracking performance on soft terrains. Notwithstanding, these studies have overlooked the impact of latency (communication delays) on system performance. Subsequent research addressed this by investigating delay effects and revealed that even small delays could not only degrade the performance but introduce additional non-passivity issues~\cite{b7a}. Additionally, they propose a Wave Variable (WV) passivity approach with derivative control to stabilize the system and enhance the command-tracking performance.

However, the proposed WV transformation with the derivative controller in \cite{b7a} leads to a more conservative nature and poor transparency. While various studies have leveraged the passivity approaches of WV and TDPA~\cite{b3,b4}. Others have utilized control techniques that include model-based approaches, model-mediated approaches, and model-free approaches~\cite{b3,b14}. Among these, model-free predictors have been widely employed in bilateral teleoperated UGV to address delay issues, due to the complexities of modeling driver behavior and the complex nonlinear dynamics of environment interactions~\cite{b8}. 

Zheng et al.~\cite{b7} introduced the first model-free predictor designed for closed network systems operating below 1 Hz, demonstrating its capacity to compensate for small delays (135 ms) and addressing the non-passivity issue in the system. It was later validated on teleoperated UGVs to show its efficacy in compensating large delays of 600 ms in real-world scenarios~\cite{b19,b19a} Building on this, they further combined the model-free predictor with a model-based approach to compensate for large delays of high-speed teleoperated UGVs~\cite{b8}. The results showed that this approach yielded more accurate predictions than either approach alone with a small delay. Nevertheless, challenges remain in dealing with the uncertainty of real network delays. Sridhar et al.~\cite{b20} modified the model-free predictor to better accommodate uncertain network delays by introducing an adaptively varying parameter that depends on delay fluctuation. Despite the enhanced robustness, the approach encountered issues with oscillatory error gain in higher frequency coupling variables~\cite{b3}. In a recent study~\cite{b21}, the model-free predictor was modified by introducing an additional degree of freedom as a potential solution to address both the uncertainties in delays and the challenges of higher oscillatory error gain. Results from a connected testbed with a delay of 175 ms demonstrate robustness to varying delays and improved performance, leading to a 30\% increase in the fidelity of the closed-loop integration compared to the original predictor. Nevertheless, despite these improvements, the technique falls short in modeling the complex nonlinear, and temporal dynamic behaviors in teleoperated UGV systems coupled with longitudinal slippage, particularly in handling large delays, such as those encountered in long-distance Earth-to-Lunar networks.

\subsection{Contributions}
In this work, we advance the recent delay compensation framework~\cite{b21} using the Recurrent Neural Network (RNN) to develop a high-fidelity closed-loop integration for teleoperated UGVs subjected to longitudinal slippage. The original contributions of this work are outlined as follows.

\begin{enumerate}
\item{We developed a novel physics-informed LSTM (PiLSTM)-based predictor framework to improve performance and effectively compensate for large network delays.} 

\item{A comprehensive performance comparison of delay compensation is conducted between the conventional predictor framework~\cite{b21} and our PiLSTM-based predictor framework in open-loop case studies.}

\item{The developed framework is validated for its effectiveness in mitigating the negative effects of significant real-network delays on the teleoperated UGVs in closed-loop scenarios. }
\end{enumerate}

The structure of the remainder of this article is as follows. In Section II, we delve into integrated system dynamics and controls for UGV bilateral teleoperation on soft terrains. Section III discusses the detailed design of the physics-informed LSTM-based predictor, including problem definition, data generation and preprocessing, model development, and performance evaluation. Section IV presents the human-in-the-loop experiment test platform and the experimental design protocol. Section V details the comparative analysis of delay compensation performance and evaluates overall teleoperation effectiveness. Finally, Section VI summarizes the findings and their significance, and outlines future research directions.

\section{System Dynamics and Controls}
\begin{figure*}[!t]
\centering
\includegraphics[width=6.7in]{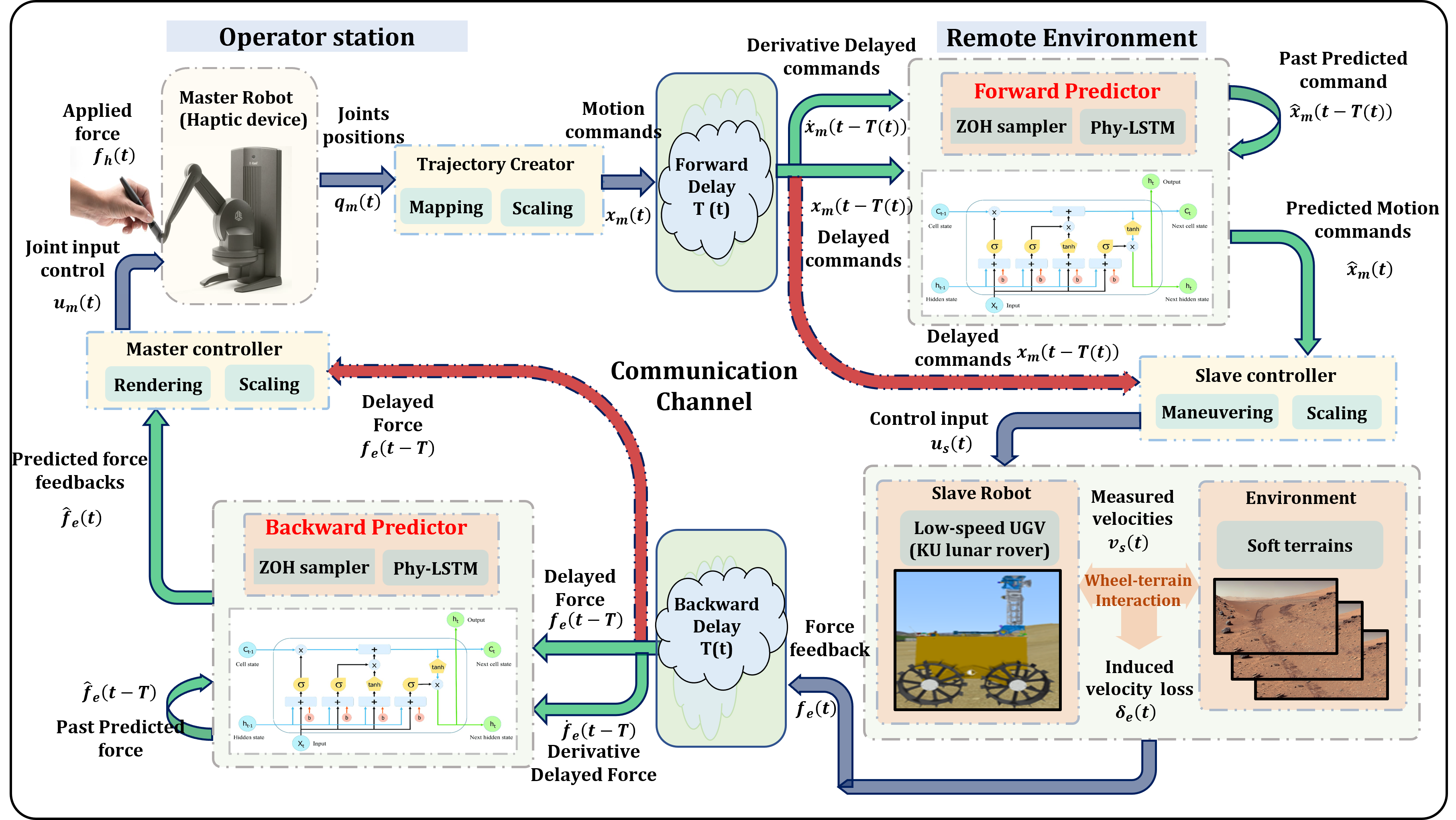}
\caption{Proposed predictor framework, comprising two model-free predictors namely; forward predictor and backward predictor, and a blended architecture for the teleoperated UGV subjected to terrain-induced slippage. The Red lines represent the Delayed case, and the Green Lines represent the Predicted case.}
\label{fig2}
\vspace{-5pt}
\end{figure*}

The core of this research lies in delay compensation for the bilateral teleoperation of low-speed UGVs subjected to induced longitudinal slippage, integrating non-holonomic motion commands and environmental force feedback to improve operator awareness of terrain-induced slippage. The architecture of the proposed predictor framework, as illustrated in Fig.~\ref{fig2}, comprises several key components that include the haptic device, the communication channel, UGV operating on soft terrain, and both forward and backward predictors.

\subsection{Haptic device (master) model}
The Phantom haptic device was utilized as the haptic interface. The two Degrees of Freedom (DOF) of the haptic device (joint 1 and joint 2) were employed to encode the linear and angular velocities of the UGV, as shown in the operator side of Fig.~\ref{fig2}. This approach is commonly used in tele-driving UGV with non-holomorphic behavior~\cite{b5b}. The nonlinear dynamic equation of the Phantom haptic device was derived in detail in~\cite{b21b}, which was eventually reduced to linear dynamics~\cite{b12} in~\eqref{firsteq}.

\begin{equation}
\mathbf{M}_m \ddot{q}_m(t) + \mathbf{C}_m \dot{q}_m(t) = u_m(t) + f_h(t)
\label{firsteq}
\end{equation}

\noindent where $\mathbf{M}_m$ represents the mass matrix of the robot, $\mathbf{C}_m$ denotes the centrifugal matrix, $q_m = \begin{bmatrix} q_{m v} & q_{m \omega} \end{bmatrix}^T$ is the joints position variable, $u_m = \begin{bmatrix} u_{m v} & u_{m \omega} \end{bmatrix}^T$ is the joints control input, and $f_h = \begin{bmatrix} f_{h v} & f_{h \omega} \end{bmatrix}^T$ represents the external force applied by the human operator.

To address potential instability resulting from workspace mismatch between the master and the slave, a new dynamic variable was introduced, \(x_{m}=\lambda \dot{q}_{m}+q_{m} \quad(0<\lambda<1)\). As a result, the robot controller can be expressed as \(u_{m}=\bar{u}_{m}+u_{m}^*\), where \(\bar{u}_{m}\) is the teleoperation controller, and \(u_{m}^*\) is the local controller, given by \(u_{m}^*=\mathbf{B}_{vi} \dot{q}_{m}+\mathbf{B}_{p} q_{m}\). The modified linear dynamics result in a first-order system as in~\eqref{eqn2}.

\begin{equation}
\label{eqn2}
\mathbf{\bar{M}}_m \dot{x}_m(t) + \mathbf{\bar{C}}_m x_m(t)= u_m(t)+f_h(t)
\end{equation}

\noindent where \(\mathbf{\bar{M}}_m= \mathbf{M}_m / \lambda\) and  \(\mathbf{\bar{C}}_m= \mathbf{C}_m / \lambda\) are the equivalent mass and centrifugal force coefficients, respectively. $x_m = \begin{bmatrix} x_{m v} & x_{m \omega} \end{bmatrix}^T$ is the new joints variable for motion commands.


\subsection{UGV (slave) Model}
For this research, the low-speed teleoperated UGV utilized was the Khalifa University Lunar rover, which is an exact replica of the UAE Rashid rover~\cite{b5aa}, as displayed in the remote environment of Fig.~\ref{fig2}. The UGV's mobility concept relied on its kinematic models that dictate the non-holomorphic motion, especially at low speeds. Inspired by the Boogie rover, the UGV locomotion principle assumes that the speeds of the front and rear wheels are nearly equal when operating on flat terrain. The kinematic model in~\eqref{eqn3} corresponds to the assumption of pure rolling, is formulated in the UGV frame~\cite{b5b}.

\begin{equation}
\label{eqn3}
\left[\begin{array}{c}
v_s(t) \\
\omega_s(t)
\end{array}\right]=\left[\begin{array}{cc}
1 / 2 & 1 / 2 \\
-1 / (2b) & 1 / (2b)
\end{array}\right]\left[\begin{array}{c}
v_r(t) \\
v_l(t)
\end{array}\right]=\left[\begin{array}{c}
u_{sv}(t) \\
u_{s\omega}(t)
\end{array}\right]
\end{equation}

\noindent where \(v_s(t)\) and \(\omega_s(t)\) represent the measured linear and angular velocity of the UGV, respectively. \(u_{sv}(t)\), and \(u_{s\omega}(t)\) represent the desired linear and angular velocity of the UGV, respectively. \(b\) is the distance between the right and the left wheels, \(v_r(t)\) and \(v_l(t)\) are the right and the left measured wheel velocities, respectively. The transformation matrix, denoted as \(\mathbf{E}(b)\), is:

\begin{equation}
\label{eqn4}
\mathbf{E}(b) =\left[\begin{array}{cc}
1 / 2 & 1 / 2 \\
-1 / (2b) & 1 / (2b)
\end{array}\right]
\end{equation}

On soft terrains, like loose soil, the relationship between the linear velocity of each wheel and the product of the wheel's angular velocity and radius may not be equal due to wheel longitudinal slippage caused by interaction forces from the terrain \cite{b6}. Here we use \eqref{eq:slippage} to describe the level of wheel slippage. This slippage deviation from the original kinematic model (in \eqref{eqn3}) was recognized by Li et al. \cite{b12}. They proposed a modified kinematic model for the UGV on soft terrains, expressed by \eqref{eqn5}.

\begin{equation}
s_r(t) = \frac{v_{rd}(t) - v_r(t)}{v_r(t)}, \ s_l(t) = \frac{v_{ld}(t) - v_l(t)}{v_l(t)} 
\label{eq:slippage}
\end{equation}

\begin{equation}
\label{eqn5}
\left[\begin{array}{c}
v_s(t) \\
\omega_s(t)
\end{array}\right]=\left[\begin{array}{c}
u_{sv}(t) \\
u_{s\omega}(t)
\end{array}\right] - \mathbf{E}(b) \left[\begin{array}{c}
v_{rd}(t) - v_r(t) \\
v_{ld}(t) - v_l(t)
\end{array}\right]
\end{equation}

Where \(v_{rd}(t)\) and \(v_{ld}(t)\) are the right and the left desired wheel velocities, respectively. \(s_{r}(t)\) and \(s_{l}(t)\) are the right and the left desired wheel slippage, respectively. Besides, we only care about the case of $v_{rd} \geq v_r$ in (5), corresponding to $s_r \geq 0$, to avoid the non-passivity issues due to negative slippage~\cite{b5b}. To solve this issue, their feedforward controller (FFC) was adopted to compensate for the induced slippage to improve the command-tracking performance in (6). Now the terrain-induced longitudinal slippage $f_e(t)$, the functional relationship between its linear and angular velocity, is defined based on the induced velocity loss of the UGV, as in~\eqref{eqn6}:

\begin{equation}
\label{eqn6}
\left[\begin{array}{c}
f_{ev}(t) \\
f_{e\omega}(t)
\end{array}\right]=\left[\begin{array}{c}
u_{sv}(t) -  v_s(t) \\
u_{s\omega}(t) - \omega_s(t)\end{array}\right]=\mathbf{E}(b) \left[\begin{array}{c}
v_{rd}(t) - v_r(t) \\
v_{ld}(t) - v_l(t)
\end{array}\right]
\end{equation}

In this study, $f_e$ is the environmental force, which is assumed the friction force exerted by the soft terrains on the wheel and the UGV has a high-precision built-in velocity controller for each wheel motor, thus the control input is $u_s$. As the induced slippage is well compensated by the FFC, the command-tracking performance in (6) becomes ideal in (8). Refer to \cite{b7a,b12} for more detailed.

\begin{equation}
\begin{bmatrix}
v_s(t) \\
\omega_s(t)
\end{bmatrix}
\overset{f_e \rightarrow 0}{\longrightarrow}
\begin{bmatrix}
u_{sv}(t) \\
u_{s\omega}(t)
\end{bmatrix}
\end{equation}

\subsection{Proposed Bilateral Teleoperator Controls}

While the existing bilateral teleoperator controllers have served their purpose in mapping coordination between the haptic device (master) and the UGV (slave), as demonstrated by the delayed case shown with red lines in Fig.~\ref{fig2}, their performance and transparency are poor in the presence of large delays $T$~\cite{b7a}. To address this shortcoming, we propose new bilateral teleoperator controllers designed to enhance performance and transparency while maintaining overall system stability. These controllers ensure effective coordination between the haptic device (master) and the UGV (slave), as demonstrated by the predicted case with green lines in Fig.~\ref{fig2}, even in the presence of large delays $T$. The control designs incorporate simple proportional derivative controllers, with the mathematical description of the transition outlined as follows:

\textbf{Master Controller:} To achieve perfect coordination between \((x_{mv}, v_s)\) and \((x_{m\omega}, \omega_s)\) at the operator side. The joint control input \(u_m\) in (2) is switched from~\eqref{eqn8} to~\eqref{eqn88} by introducing predicted feedback (\(\hat{f}_{ev}(t)\), \(\hat{f}_{e\omega}(t)\)):
    
    \begin{equation}
    \label{eqn8}
    \left[\begin{array}{l}
    u_{mv}(t)\\
    u_{m\omega}(t)
    \end{array}\right] = \left[\begin{array}{l}
    -k_{mv} \ f_{ev} (t-T)\\ 
    -k_{m\omega} \ f_{e\omega}(t-T) 
    \end{array}\right]
    \end{equation}

    \begin{equation}
    \label{eqn88}
    \left[\begin{array}{l}
    u_{mv}(t)\\
    u_{m\omega}(t)
    \end{array}\right] = \left[\begin{array}{l}
    -k_{mm} \ \hat{f}_{ev} (t) \\ 
    -k_{m\omega} \ \hat{f}_{e\omega}(t) 
    \end{array}\right]
    \end{equation}

\textbf{Slave Controller:}  To achieve perfect coordination between \((x_{mv}, v_s)\) and \((x_{m\omega}, \omega_s)\) at the remote side. The control input \(u_s\) in (6) is switched from~\eqref{eqn9} to~\eqref{eqn99} by introducing predicted motion commands (\(\hat{x}_{mv}(t)\), \(\hat{x}_{m\omega}(t)\)):
    
    \begin{equation}
    \label{eqn9}
    \left[\begin{array}{l}
     u_{sv}(t)\\
     u_{s\omega}(t)
    \end{array}\right] = \left[\begin{array}{l}
     k_{sv} \ x_{mv}(t-T) \\
     k_{s\omega} \ x_{m\omega}(t-T)
    \end{array}\right]
    \end{equation}

    \begin{equation}
    \label{eqn99}
    \left[\begin{array}{l}
     u_{sv}(t)\\
     u_{s\omega}(t)
    \end{array}\right] = \left[\begin{array}{l}
     k_{sv} \ \hat{x}_{m1}(t) \\
     k_{s\omega} \ \hat{x}_{m2}(t)
    \end{array}\right]
    \end{equation}

The proportional control gains \(k_{mv}\), \(k_{m\omega}\), \(k_{sv}\), and \(k_{s\omega}\) are maintained positive~\cite{b10}. Therefore, in the designed teleoperation system above, if the desired linear velocity \(x_{mv}(t)\) is greater than the actual linear velocity \(v_s(t)\), a backward force will be felt by the human operator that pushes back on the master robot joint 1. Similarly, for any difference in angular velocities \(x_{m\omega}(t)\) and \(\omega_s(t)\), a backward force will be felt on joint 2. Conversely, if the desired linear velocity \(x_{mv}(t)\) is smaller than the actual linear velocity \(v_s(t)\), a forward force will be felt by the human operator that pulls the master robot joint 1 forward. Similarly, for any difference in angular velocities \(x_{m\omega}(t)\) and \(\omega_s(t)\), a backward force will be felt on joint 2. Thus, the force feedback serves as a slippage awareness guide to the human operator, enabling more effective command execution, as demonstrated in~\cite{b7a,b10,b12}.


\section{Predictor framework Design}
\label{sec:Predictor}
The fundamental form of the conventional model-free predictors adopted in this paper was initially developed to compensate for delays in general networked closed-loop systems~\cite{b7}, then it was later expanded for the high-speed teleoperated UGV in~\cite{b8}. In~\cite{b21}, the modified predictor has been developed based on the first-order plus time delay system dynamics, as described in~\eqref{eqn11}. 
\begin{equation}
\begin{aligned}
\label{eqn11}
\dot{x}_p(t) &= \dot{x}(t-T) + \beta\left[x(t-T) - x_p(t-T)\right] \\
&\quad + \alpha\left[\dot{x}(t-T) - \dot{x}_p(t-T)\right], \\
\hat{x}(t) &= x_p(t).
\end{aligned}
\end{equation}

The predictor equation, \eqref{eqn11}, incorporates the actual state $x(t-T)$ and its derivative $\dot{x}(t-T)$ at a previous time step $t-T$; the predicted state $x_p(t-T)$ and its derivative $\dot{x}_p(t-T)$ at the same previous time step $t-T$; and the predicted state $x_p(t)$ at the current time step $t$, to estimate the predicted undelayed state $\hat{x}(t)$, which is an approximation of the actual undelayed state $x(t)$. The performance of the predictor is adjusted using the regularization parameters $\beta$ and $\alpha$ for a specific delay $T$. Details of the predictor stability analysis and performance can be found in~\cite{b21}.

The primary limitation of the above conventional predictor dynamics lies in their failure to capture the complex, temporal, and unknown dynamics behaviors of the coupling variable, significantly compromising the prediction performance~\cite{b21}. A promising solution to address this limitation is the implementation of Recurrent Neural Network (RNN)-based predictors. These predictors, as proven in~\cite{b22,b23}, are effective at learning nonlinear and temporal dynamics behaviors, thus making them ideal tools for modeling highly nonlinear and time-dependent systems. However, they often struggle with generalization in complex dynamic systems~\cite{b25,b27}. This challenge paves the way for hybrid models that combine data-driven insights with some physical constraints~\cite{b28,b29}. Capitalizing on this, our study introduces a new RNN-based predictor framework, fundamentally structured based on the Long Short-Term Memory (LSTM) networks, augmented with some physical constraints, similar to the architecture in~\cite{b29}. The proposed framework consists of two distinct predictors: a forward predictor tailored to predict motion commands and a backward predictor tailored to force feedback, as in Fig.~\ref{fig2}.

\subsection{Problem Definition}
The proposed predictor is utilized to estimate the undelayed variables from delayed variables. The focus of the models is to perform a single-step prediction of the coupling variables, particularly in the context of a track-following task. Given the nature of the task, it is imperative to accurately predict future variables at a time step.

The problem can be defined as:
\begin{equation}
\label{eqn13}
  \mathcal{H}_{\Theta}(X_t, X_{t-1}, X_{t-2}, \ldots, X_{t-n}) = Y_{t+1},
\end{equation}
where $\mathcal{H}_{\Theta}$ denotes the PiLSTM network with parameters $\Theta$, mapping the sequence inputs $(X_t, X_{t-1}, X_{t-2}, \ldots, X_{t-n})$ to a single output $Y_{t+1}$ at time step $t$ during the learning process. Each input vector is defined as $X = [x(t - T), x_p(t - T), \dot{x}(t - T), \dot{x}_p(t - T)]$ and corresponds to the one-step ahead actual output $Y = x(t)$.

\subsection{Data and Preproccesing}
The dataset was generated from the human-in-the-loop experiment of the teleoperated UGV on soft terrains, which is discussed in detail in Section~\ref{sec:Experiment}. The dataset acquired different path patterns. We initiated the process by gathering comprehensive data of motion commands $x_m(t)$ and the force feedback signals $f_e(t)$, saved at a frequency of 10 Hz from the experiments. The data comprise five essential variables for each of the coupling variables, including the undelayed actual value $x(t)$, delayed actual value $x(t-T)$, the delayed predicted value $x_p(t-T)$, the derivative of the delayed actual value $\dot{x}(t-T)$, and the derivative of the delayed predicted value $\dot{x}_p(t-T)$. These variables are assumed to capture the essential information necessary for accurately modeling the complexity of the empirical data. 

The training dataset was recorded from three randomly selected trained operators with a sampling rate of 10 Hz for a total simulation time of 300 s (comprising about 3000 acquisitions). For the testing datasets, three sets were collected from different trained operators, with each testing dataset maintaining the same sampling rate as the training set but spanning a shorter simulation duration of 30 s. Furthermore, the training dataset is further partitioned into two distinct segments for training and validation purposes, in the ratio of 70\% to 30\%, respectively. Each segment is structured with the input features $x(t-T)$, $x_p(t-T)$, $\dot{x}(t-T)$, and $\dot{x}_p(t-T)$; and the target output feature $x(t)$, which represents the ground truth values that the model aims to learn and predict. Additionally, the input features were preprocessed by normalizing and scaling the features to ensure homogeneity among the features, similar to the approach described in~\cite{b23}.

\subsection{Model Development}
Here, we discuss the detailed design of the four distinct models, each tailored to one of the four coupling variables of the bilateral teleoperation system. These include the desired linear velocity \(x_{mv}(t)\), desired angular velocities \(x_{m\omega}(t)\), longitudinal force feedback \(f_{ev}(t)\), and lateral force feedback \(f_{e\omega}\).

\subsubsection{Network Architecture}
\begin{figure}[htbp]
\centering
\includegraphics[width=\columnwidth]{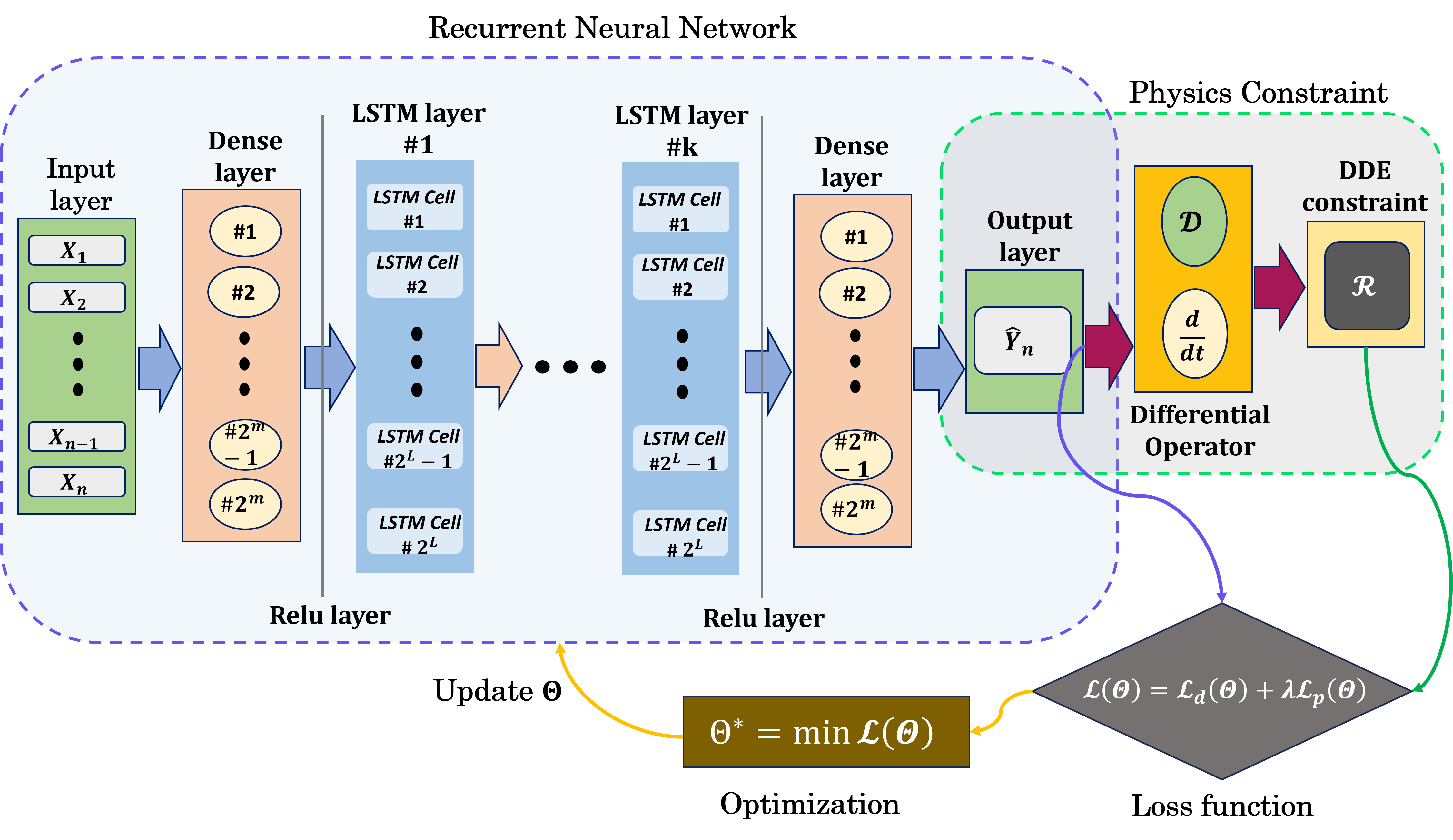}
\caption{Physics-informed LSTM architecture for the proposed predictor: Number of fixed-length input features \(n\), number of dense layer hidden units \(m\), number of LSTM layers \(k\), number of LSTM layer hidden units \(l\), and a single-step target output at \(n\).}
\label{fignetwork}
\vspace{-5pt}
\end{figure}

The architecture of our physics-informed LSTM network is designed to handle sequential data with complex nonlinearity and temporal dependencies \cite{b29}. The network is composed of an input layer, a dense layer, an activation layer, LSTM layers, and an output layer, along with a physics constraint, as depicted in Fig. \ref{fignetwork}. Each LSTM layer captures different levels of abstraction within the sequence data.

The input layer receives the input sequence, composed of a feature vector of fixed length, denoted as \((X_1, X_2, \ldots, X_n)\), where each \(X_i\) is defined as \(X_i = [x(t_i - T), x_{p}(t_i - T), \dot{x}(t_i - T), \dot{x}_{p}(t_i - T)]\) for \(i = 1, \ldots, n\). An LSTM layer is comprised of a series of LSTM cells, each employing a sigmoid activation function \(\sigma\) for gate operations and a hyperbolic tangent \(\tanh\) for updating cell states. Its unique feedback mechanism, where the output of each cell is fed back into itself at the next time step, endows the network with the capability to capture long-term temporal dependencies from the input sequence. The output layer produces single-step predicted values, represented by \(\hat{Y}_n\), where \(\hat{Y} = \hat{x}(t_i) = x_p\). Thus, each of our four models shares the same architectural structure but may vary in the number of layers and units according to their complexity.

\subsubsection{Loss function formulation}
Having discussed how to design the network architecture, we next address how to obtain the network parameters \(\Theta\). For this, the optimization problem is defined as follows:
\begin{equation}
\label{eqn14}
\Theta^* = \min \mathcal{L}(\Theta)
\vspace{-5pt}
\end{equation}
where \(\mathcal{L}\) represents a custom loss function, which must be differentiable with respect to its parameters. In neural networks addressing this kind of regression problem, the mean absolute error is the most commonly used for data loss as given by:
\begin{equation}
\label{eqn15}
\mathcal{L}_d(\Theta) = \frac{1}{N} \sum_{i=1}^{N} \left| x_p(t_i) - x(t_i) \right|
\vspace{-5pt}
\end{equation}
where \(N\) represents the total number of samples in a run. This loss function is selected due to its inherent benefit of reducing the impact of outliers. To further enhance model generalization to unseen data~\cite{b29}, some physical constraints are enforced on the derivative of the variable. By employing a method in~\cite{b24} to approximate the temporal derivative of past states on the right-hand side of the first-order dynamics in (13), it is transformed into a standard Delay Differential Equation (DDE). The compacted form of this DDE is described in~\eqref{eqn15b}, with \(\psi(t) = t - T\). In turn, the loss function assimilates the DDE constraint, which introduces a soft regularization in the optimization process to ensure compliance with fundamental principles. The physics-informed loss is defined based on the L2-norm of the DDE residual, as presented in~\eqref{eqn16}.
\begin{equation}
\label{eqn15b}
\begin{aligned}
\dot{x}_p(t) &= \mathcal{F} (t, x(\psi(t)), x_p(\psi(t)) \\
\hat{x}(t) &= x_p(t)
\end{aligned}
\end{equation}
\begin{equation}
\label{eqn16}
\mathcal{L}_p(\Theta) = \frac{1}{M} \sum_{j=1}^{M} \left\| \dot{x}_p(t_j) -  \mathcal{F} (t_j, x(\psi(t_j)), x_p(\psi(t_j)) \right\|^2
\vspace{-5pt}
\end{equation}
where \(M\) represents the number of collocation points where the DDE constraint is enforced. The overall cost function \(\mathcal{L}(\Theta)\) is a weighted sum of these two loss terms and is minimized with respect to the network parameters \(\Theta\) during the training process. By optimizing this loss, the network learns a model \(\mathcal{H}_{\Theta}\) that not only fits the observed data but also aligns with the specified dynamics of the system.
\begin{equation}
\label{eqn17}
\mathcal{L}(\Theta) = \mathcal{L}_d(\Theta) + \lambda\mathcal{L}_p(\Theta)
\end{equation}
The coefficient \(\lambda\) is a hyperparameter that balances the two loss components~\cite{b29}, allowing you to adjust the relative importance of data fitting and physical constraint adherence during training as depicted in Fig.~\ref{fignetwork}.

\subsubsection{Training}

The training protocol for each of the models is initiated by inputting a sequence of 50 previous data points for each model (5 s of historical data). This predetermined sequence length, denoted as $n$, serves as the input vector for the network. The network parameters undergo iterative refinement through an optimization routine predicated on a composite loss function, as delineated in~\eqref{eqn17}. The backpropagation through time (BPTT) method is applied, and gradient descent algorithms are also employed using the Adam optimizer for its proficiency in stable convergence and adaptive learning rate adjustments. The pseudo-code of the training process, encapsulated in Algorithm~\ref{alg:alg1}, delineates the computational steps undertaken during the model learning phase.
\begin{algorithm}
\caption{Physics-Informed LSTM Training Pseudo Code}\label{alg:alg1}
\begin{algorithmic}
\STATE {\textbf{INPUT:} $X = [X_1, X_2, \ldots, X_n]$, each $X_i \in \mathbb{R}^n$}
\STATE {\textbf{OUTPUT:} $Y = Y_n$, where $Y_n \in \mathbb{R}^n$}
\STATE {\textbf{PARAMETERS:} 
$\Theta$ = [$W_{in}$, $b_{in}$, $W_f$, $U_f$, $b_f$, $W_{\tilde{c}}$, $U_{\tilde{c}}$, $b_{\tilde{c}}$, $W_i$, $U_i$, $b_i$, $W_o$, $U_o$, $b_o$, $W_{out}$, $b_{out}$]}
\STATE {\textbf{INITIALIZE:} hidden state $h_0 = \vec{0}$, cell state $c_0 = \vec{0}$, learning rate $\alpha$, physics loss weight $\lambda$}
\STATE {\textbf{DEFINE:} differential operator $\mathcal{D}$, DDE constraint $\mathcal{R}$}
\STATE \textbf{TRAIN and VALIDATE}$(X_{train}, Y_{train},X_{val}, Y_{val}, \mathcal{D}, \mathcal{R})$
\STATE \hspace{0.3cm} \textbf{for each} epoch \textbf{do}
\STATE \hspace{0.7cm} \textbf{for each} batch $(X_{batch}, Y_{batch})$ \textbf{in} $(X_{train},Y_{train})$ \textbf{do}
\STATE \hspace{1.1cm} Compute LSTM outputs for batch: $H_{batch}$
\STATE \hspace{1.1cm} Compute Dense output for batch: $\hat{Y}_{batch}$
\STATE \hspace{1.1cm} Apply $\mathcal{D}$ to get $\dot{\hat{Y}}_{batch}$
\STATE \hspace{1.1cm} Compute data loss: $\mathcal{L}_d(\Theta) = f(\hat{Y}_{batch}, Y_{batch})$
\STATE \hspace{1.1cm} Compute phy. loss: $\mathcal{L}_p(\Theta) = f(\dot{\hat{Y}}_{batch}, \mathcal{R}(Y_{batch}))$
\STATE \hspace{1.1cm} Compute total loss: $\mathcal{L}(\Theta) = \mathcal{L}_d(\Theta) + \lambda \mathcal{L}_p(\Theta)$
\STATE \hspace{1.1cm} Update $\Theta$ using gradient descent: 
\STATE \hspace{1.9cm} $\Theta = \Theta - \alpha \nabla_{\Theta} \mathcal{L}(\Theta)$
\STATE \hspace{0.7cm} \textbf{end for}
\STATE \hspace{0.7cm} Take $(X_{val}, Y_{val})$
\STATE \hspace{0.7cm} Compute output for validation: $\hat{Y}_{val}$
\STATE \hspace{0.7cm} Compute validation loss: $\left\lVert \hat{Y}_{val} - Y_{val} \right\rVert_2$
\STATE \hspace{0.3cm} \textbf{end for}
\STATE \textbf{CHECK POINT}: $\Theta^* = \Theta_{best}$
\end{algorithmic}
\end{algorithm}
\subsubsection{Hyperparameter tuning}
Bayesian optimization is employed to obtain the optimal hyperparameters for the network models. Table~I summarizes the optimal hyperparameters as obtained for each of the four distinct models.
\begin{table}
\centering
\caption{Optimal Hyperparameters for Predictors}
\label{tab:hyperparameters}
\begin{tabular}{|p{1.2cm}|p{0.95cm}|p{1.1cm}|p{0.9cm}|p{1cm}|p{1cm}|}
\hline
\textbf{Coupling Variables} & \textbf{Initial Learn Rate} & \textbf{Dense Hidden Units, $m$} & \textbf{LSTM Depth, $k$} & \textbf{LSTM Hidden Units, $l$} & \textbf{Threshold} \\
\hline
$x_{mv}$ & 0.00230 & 118 & 2 & 162 & 0.07 \\
\hline
$x_{m\omega}$ & 0.00016 & 174 & 3 & 92 & 1.12 \\
\hline
$f_{ev}$ & 0.00095 & 151 & 2 & 188 & 0.80 \\
\hline
$f_{e\omega}$ & 0.00076 & 207 & 4 & 75 & 0.92 \\
\hline
\end{tabular}
\vspace{-10pt}
\end{table}
\subsection{Prediction Performance}
During the training process, our methodology entailed a comprehensive evaluation of various RNN layers that include LSTM, GRU, and hybrid LSTM-GRU to have the best models. For rigorous model validation, we utilized testing datasets that were distinct from those employed during the training phase, to prove the model generalization and their efficacy in extrapolating insights from new, unseen data. Thus, to prevent overfitting the model with the training data, the performance of the developed models is quantified using the Root Mean Square Error (RMSE), mathematically described as follows:
\begin{equation}
RMSE = \sqrt{\frac{1}{N} \sum_{i=1}^{N}(x_p(t_i)-x(t_i))^2}
\end{equation}
where \( N \) represents the total number of samples within the dataset. The evaluation was conducted in an offline mode, enabling a thorough analysis of the models' capabilities without the constraints of real-time processing. This approach allowed for a quick and detailed evaluation of the models' performance. Table~II provides the performance analysis of the final developed models.

\begin{table}[ht]
\centering
\caption{Performance Metrics for Different RNN Network Architectures}
\begin{tabular}{|p{1.2cm}|p{1.4cm}|p{0.9cm}|p{0.9cm}|p{0.9cm}|p{0.9cm}|}
\hline
\textbf{Coupling Variables} & \textbf{Network Arch.} & \textbf{Train RMSE} & \textbf{Test \#1 RMSE} & \textbf{Test \#2 RMSE} & \textbf{Test \#3 RMSE} \\
& & \textbf{x\(10^{-3}\)} & \textbf{x\(10^{-3}\)} & \textbf{x\(10^{-3}\)} & \textbf{x\(10^{-3}\)} \\
\hline
\(x_{mv}\) & LSTM & 2.90 & 3.71 & 3.26{\cellcolor{greenCell}} & 4.05 \\
& GRU & 3.31 & 4.49 & 4.04 & 4.92{\cellcolor{yellowCell}} \\
& LSTM+GRU & 3.31 & 4.36 & 4.00 & 4.84 \\
\hline
\(x_{m\omega}\) & LSTM & 50.7 & 67.9 & 62.1{\cellcolor{greenCell}} & 72.3 \\
& GRU & 52.2 & 76.1 & 75.2 & 81.1{\cellcolor{yellowCell}} \\
& LSTM+GRU & 51.9 & 75.0 & 70.9 & 72.6 \\
\hline
\(f_{ev}\) & LSTM & 15.7 & 21.4 & 19.9{\cellcolor{greenCell}} & 31.2{\cellcolor{yellowCell}} \\
& GRU & 15.2 & 22.3 & 20.8 & 25.6 \\
& LSTM+GRU & 15.3 & 21.2 & 24.9 & 29.0 \\
\hline
\(f_{e\omega}\) & LSTM & 36.1 & 57.9 & 52.1{\cellcolor{greenCell}} & 61.3 \\
& GRU & 37.4 & 59.1 & 55.2 & 68.1{\cellcolor{yellowCell}} \\
& LSTM+GRU & 37.0 & 59.0 & 54.9 & 66.7 \\
\hline
\end{tabular}
\vspace{-5pt}
\end{table}

The analysis of the error values from Table~II indicates that the average generalization error for \(x_m\) (i.e., 26\% for the LSTM) is consistently lower than those for \(f_e\) (i.e., 52\% for the LSTM), which might suggest that the models are better at predicting motion commands than force feedback. This disparity in prediction accuracy may be attributed to the inherent nature of the data, where motion commands exhibit a more regular and predictable pattern compared to the more complex and less predictable patterns of force feedback. In terms of network architecture, the LSTM consistently outperforms the GRU and the combined LSTM+GRU in predicting both \(x_m\) and \(f_e\), as evidenced by the lowest error values highlighted in green. Moreover, the LSTM seems to show the least variation in error rates across the three test datasets, suggesting better robustness to varying conditions compared to the other two. Conversely, the GRU exhibits the highest error values, which are highlighted in brown. Furthermore, Test 2 yields the most accurate predictions for all variables with the least average generalization error, while Test 3 consistently shows the least accurate performance. This pattern suggests that the closeness of the test data to the training data could be a determining factor in the prediction performance of the models. Finally, the best models (least errors) of the four coupling variables are chosen. 

\section{Human-in-the-loop Experiments}
\label{sec:Experiment}
Human-in-the-loop experiments as in~\cite{b7a,b19a,b30} were performed on a test platform to evaluate the performance of teleoperated UGVs on soft terrain with and without the predictor framework in the presence of large delays.

\subsection{Test Platform}
\begin{figure}[htbp]
  \centering
  \includegraphics[width=3.4in]{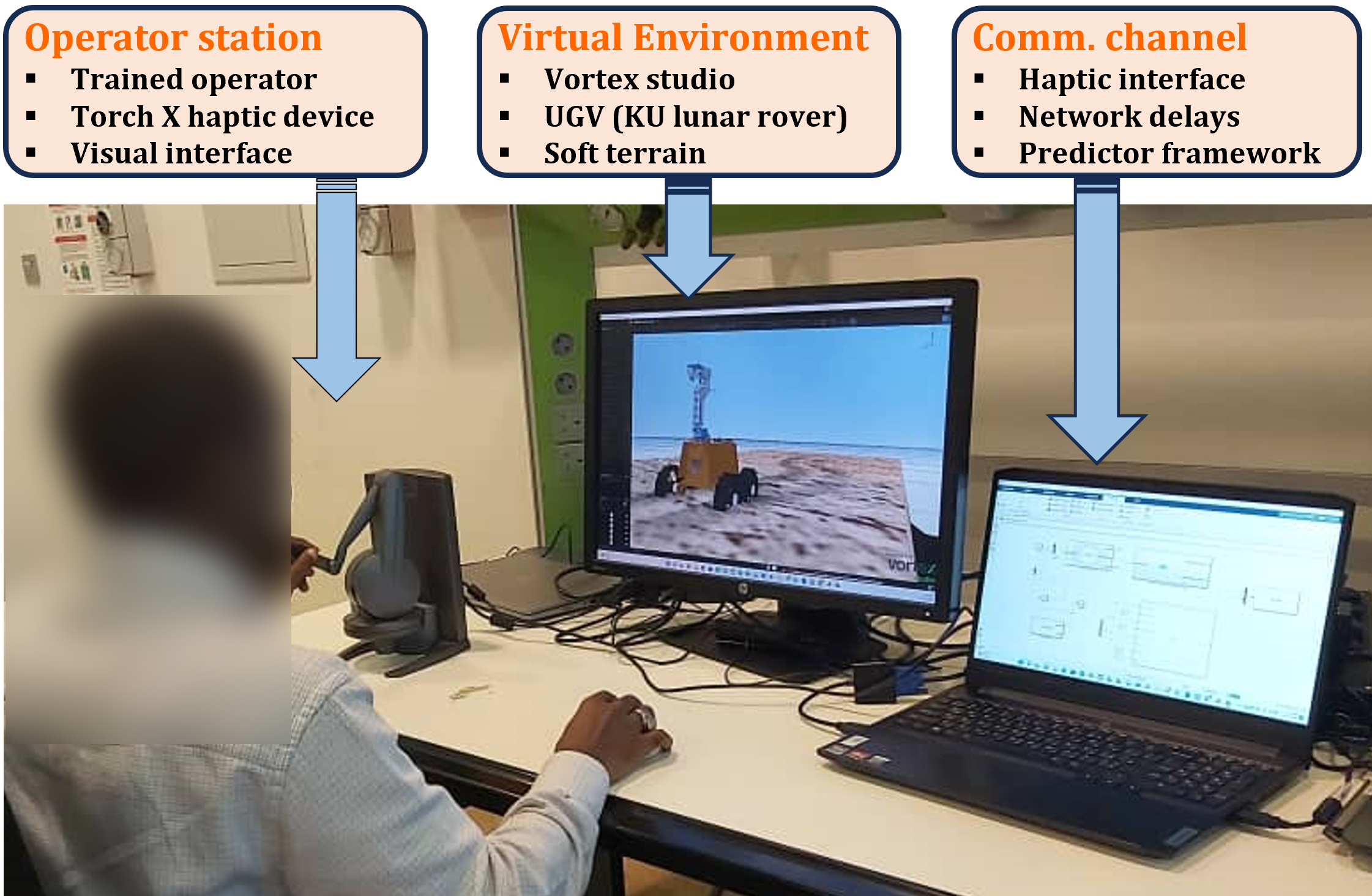}
  \caption{Human-in-the-loop test platform for teleoperated UGV traversing soft terrain with slippage awareness.}
  \label{operatorpic}
  \vspace{-5pt}
\end{figure}

A real-time human-in-the-loop test platform has been developed, integrating a Torch-X haptic device with MATLAB/Simulink and the Vortex Studio simulator to emulate a low-speed teleoperated UGV, as shown in Fig.~\ref{operatorpic}. The choice of Vortex Studio was based on previous work~\cite{b10,b12}, which demonstrated its high-fidelity and realistic simulation capabilities for UGVs operating on soft terrains. Additionally, the dynamic wheel-terrain interactions were experimentally validated in~\cite{b26}. This setup can reproduce real-world longitudinal slippage of the UGVs on soft terrains.

The operator station is equipped with a Torch-X haptic device to generate motion commands and perceive force feedback signals, as well as a monitor to provide visual feedback from the virtual environment. The device, which has three joints, activated the first two for operation while the third is looked with high gain to encode non-holomorphic driving commands. The haptic device is connected as a control interface to the simulator via Simulink. To regulate the observed high-frequency commands and feedback, a first-order low-pass filter with a cutoff frequency of 0.8 Hz is applied to the coupling variables to prevent exceeding the predictor framework's bandwidth of 1.0 Hz.

The virtual environment, assumed as the remote environment, receives motion commands from the operator station to control the UGV. This UGV is simulated in Vortex Studio with a comprehensive model that includes a kinematic model, sensor suite, a Reece tire model, and a deformable terrain model, providing a dynamic representation of the UGV within the virtual environment. The deformable terrain property is fine-tuned using empirical lunar soil parameters from the Mohammed Bin Rashid Space Centre in~\cite{b26}, ensuring the simulated terrain closely mirrors the mechanical properties of lunar soil. Moreover, sensors measure the environmental forces in~\eqref{eqn6}, which are transmitted back to the operator station to provide real-time slippage awareness.

\subsection{Experiment Design}
\begin{figure}[!ht]
  \centering
  \includegraphics[width=\columnwidth]{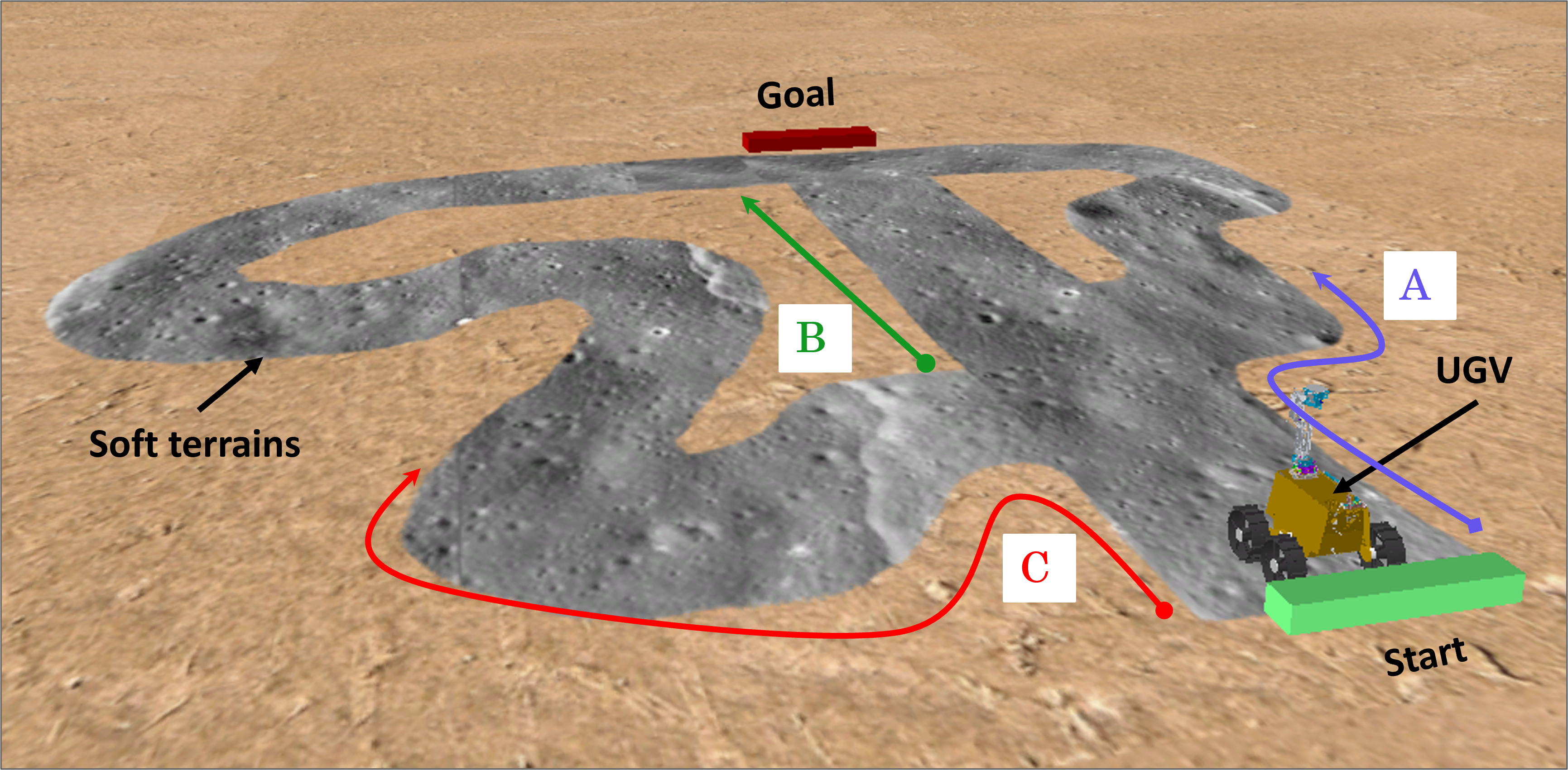}
  \caption{Designated soft terrain tracks A, B, and C; UGV, and non-track zone in the virtual environment.}
  \label{trackspicA}
  \vspace{-5pt}
\end{figure}

A roadmap was established to ensure a diverse array of test tracks, generated in a virtual environment with the lunar environment in mind, as depicted in Fig.~\ref{trackspicA}. Three distinct track patterns, labeled A, B, and C (in gray), are designed as soft terrains, starting from the green mark and ending at the red mark. Track A is a straight pattern, measuring \(10\) meters in length and \(2\) meters in width. Track B features right turns, while Track C includes both right and left turns. Each track is designed to exhibit position-dependent variable slippage, characterized by an internal friction angle, \( \phi(z) \), which varies as a function of the track's position, \(z\). This function is defined within the range \(0.50 \, \text{rad} \leq \phi(z) \leq 0.95 \, \text{rad}\), to introduce potential sand traps~\cite{b7a}. The non-track area (in brown) serves as a landmark, giving operators a sense of the distance to the turns and the speed. The safe speed on the tracks is predetermined as a maximum allowable speed of \(0.1 \, \text{m/s}\), allowing the UGV to navigate soft terrain without tipping over. The UGV parameters can be found in~\cite{b26}.

Furthermore, the one-way communication delay from the operator station to the virtual environment is assumed to be 1.25 s, representing the latency for the long earth-to-lunar distance (380,000 km) network. We employ varying delays obtained on a real network of $1.0 \sim \mathcal{U}(-0.25, 0.25)$ using the User Datagram Protocol to simulate both forward and backward delays for such long distances. The visual interface simulates lunar low-light conditions, about 0.5 lux, while the simulation editor is set to a lunar gravity of \(1.633 \, \text{m/s}^2\).

Three distinct cases are introduced under different conditions: (I) \textbf{Ideal case}, without any delays, serving as a baseline for performance; (II) \textbf{Delayed case}, with round-trip delays and without a predictor-based framework, to assess the impact of delays on performance; and (III) \textbf{Predicted case}, with the same amount of delays and the predictor framework, to assess performance improvement due to delay compensation. To conduct these experiments, five operators were selected to perform a track-following task. Before data collection, these operators received training to familiarize them with the test platform and practice operating under different conditions—with and without delays, and with and without the predictor framework. After the training, each operator repeated each of the three experimental cases, aiming to complete the track as fast as possible, with excellent command-tracking performance and without being entrapped. A portion of the collected data on coupling variables has been allocated for the training and testing phases of the predictor models in Section~\ref{sec:Predictor}.

\subsection{Predictor Framework implementation}
Each PiLSTM predictor, outlined in Section~\ref{sec:Predictor}, is tailored to a specific coupling variable. The predictors $\hat{x}_{mv}$ and $\hat{x}_{m\omega}$ are implemented on the remote environment as the forward predictor, and $\hat{f}_{ev}$ and $\hat{f}_{e\omega}$ are on the operator station as the backward predictor, integrated to form the framework shown in Fig.~\ref{fig2}. This framework is designed for optimal delay compensation in bilateral teleoperated UGVs subjected to terrain-induced slippage~\cite{b7a}. However, seeking such optimal performance is beyond the scope of this particular application. Finally, the developed predictor framework and the blended architecture for the teleoperated UGV are evaluated in two case studies discussed in subsequent sections.

\subsection{Evaluation Metrics}
The experiment, constrained by operating the teleoperated UGV under ideal, predicted, and delayed conditions, is evaluated through two case studies, each with its evaluation metrics. The first case study, an open-loop scenario, enables comparative analysis of all conditions simultaneously~\cite{b8}, while the second, a closed-loop scenario, allows for their individual examination~\cite{b12}. Evaluation metrics are as follows.

\subsubsection{Delay compensation metrics}
In the open-loop case study, the predictors' delay compensation performance is evaluated. The output of the ideal case is considered the reference for the other two cases. The compensation performance in the time domain is quantified by normalized performance metrics, in percentage, denoted as \(\Delta_n\)~\cite{b21}. This metric is computed as the ratio of the L2 norm of the difference between the predicted output and the ideal output to the L2 norm of the difference between the delayed output and the ideal output as:
\begin{equation}
\Delta_n := \frac{\lVert \hat{x}_{pred}(t) - x_{ideal}(t) \rVert_2}{\lVert x_{del}(t) - x_{ideal}(t) \rVert_2} \times 100\%
\end{equation}
where \(x\) symbolizes the coupling variable of interest. A \(\Delta_n = 0\%\) indicates that the predictor perfectly compensates for the delays, achieving an ideal performance. A \(\Delta_n < 100\%\) suggests that the predictor compensates the delay to some meaningful amount, with its proximity to zero reflecting the extent of compensation. Conversely, \(\Delta_n > 100\%\) indicates that the predictor amplifies the negative effect of the delay.

\subsubsection{Teleoperation metrics}
In the closed-loop case study, the teleoperation performance of the predictor framework is evaluated. The ideal case is considered as the baseline for the other two cases. Three metrics quantify the teleoperation performance \cite{b12}: performance, transparency, and track completion time. The performance-velocity tracking \(\Omega\) is quantified by the deviations between the commanded motion and the UGV's actual measured states. Transparency-force feedback tracking \(\Gamma\), in contrast, is quantified by the deviation between what the operator perceives and the environmental force-induced slippage. Track completion time captures how quickly the UGV is operated along the track from the start point to the end point. The first two metrics directly reflect the fidelity of the closed-loop integration, where smaller values lead to higher fidelity, and larger values lead to lower fidelity. The mathematical expressions for these metrics are~\cite{b7a}:
\begin{equation}
\Omega := \| x_{m}(t) - v_{s}(t) \|_2, \quad \Gamma := \| f_{h}(t) - f_{e}(t) \|_2
\end{equation}

However, the force applied by the operator \(f_h\) cannot be directly measured. Thus, an estimation technique based on the system dynamics~\cite{b6,b7a,b10,b12} is utilized as:
\begin{equation}
f_h = \bar{M}_m \dot{x}_m + \bar{C}_m x_m - u_m
\end{equation}

\section{Result and Discussion}
The proposed teleoperator controller gains in \eqref{eqn88} and \eqref{eqn99} were set to $k_{m} = 5$ and $k_{s}= 1$, to achieve a balanced trade-off between performance and transparency. The optimal parameter values for the conventional predictors in (13) are $\alpha=0.57, \beta=1.12$ for the forward predictor, and $\alpha=0.64, \beta=0.91$ for the backward predictor, as they exhibit different bandwidths.

\subsection{Case Study 1}
This study compares the delay compensation performance between the developed PiLSTM predictor and the conventional (Conv) predictor frameworks in open-loop scenarios \cite{b21}.

\begin{figure}[!ht]
  \centering
  \includegraphics[width=\columnwidth]{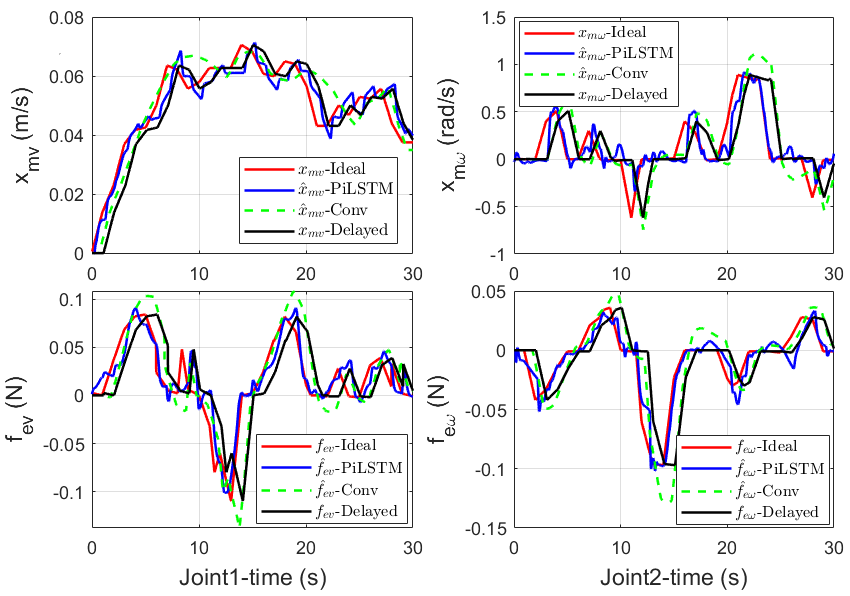}
  \caption{Operator \#1: Performance comparison of motion commands (top) and force feedbacks (bottom) among the ideal case, delayed case (1.25 s), conventional predicted case, and PiLSTM predicted case in an open-loop scenario. Our PiLSTM Predictor leads to better delay compensation performance compared to the conventional predictor.}
  \label{perm1}
  \vspace{-10pt}
\end{figure}

\begin{figure}[!ht]
  \centering
  \includegraphics[width=\columnwidth]{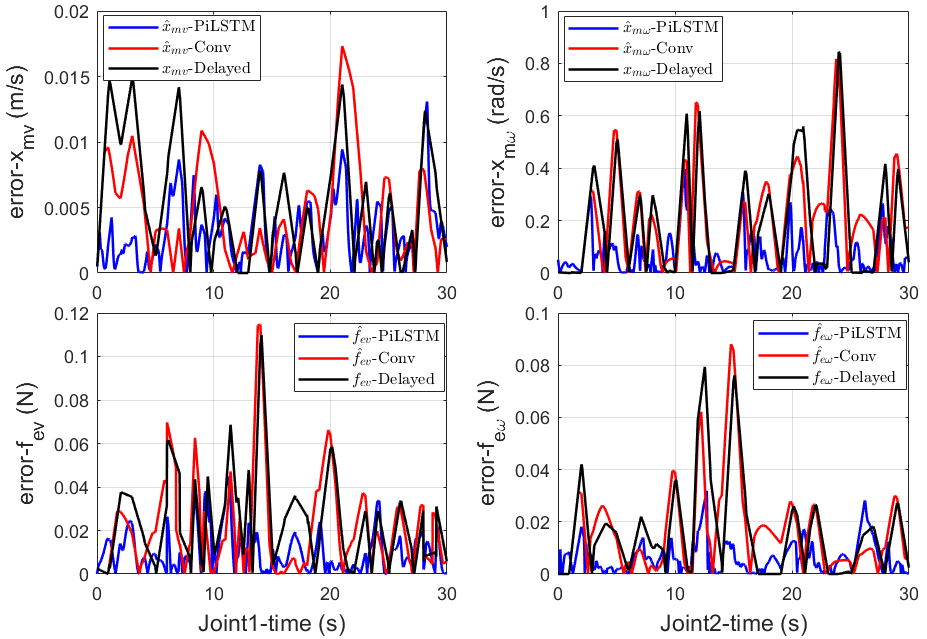}
  \caption{Operator \#1: Absolute prediction error comparison of motion commands (top) and force feedbacks (bottom) among the ideal case, delayed case (1.25 s), conventional predicted case, and PiLSTM predicted case. Limitations on the delay compensation performance of both predictors occur at specific instances, still, the conventional predictor yields relatively poor delay compensation performance.}
  \label{trackspic}
  \vspace{-5pt}
\end{figure}

A comparative analysis of the coupling variables for Operator \#1 is depicted in Fig.~\ref{perm1}. The predicted outputs from both the PiLSTM and the Conv are aligned with the Ideal output to some extent. However, the PiLSTM outputs are closer to the ideal outputs than those of the Conv, indicating better delay compensation by PiLSTM. Notably, the PiLSTM demonstrated a smoother curve, while the Conv shows more fluctuation, as evidenced by the curves. These performance differences are clearly observed in the corresponding error metrics presented in Fig.~\ref{trackspic}. It can be observed that the PiLSTM maintains a lower error margin compared to the Conv, with fewer and lower spikes throughout the observed period. In contrast, the Conv exhibits higher peaks at specific instances, which could translate to less accurate force feedback to the operator. This results in errors higher than the delayed outputs, for instance, in the Conv predicted motion commands $\hat{x}_{mv}$ and $\hat{x}_{m\omega}$ at 8.2 s and at 23.5 s, respectively, and for the force feedbacks $\hat{f}_{mv}$ and $\hat{f}_{m\omega}$ at 5.1 s and 15.8 s, respectively. These peaks typically occur in response to abrupt changes in the data, as discussed in~\cite{b21}. Further encapsulating the delay compensation performance analysis, the normalized performance metrics are summarized in Table~\ref{table3}. Evidently, the PiLSTM predicted output $\hat{x}_{mv}$ demonstrates a significantly lower error of 38.3\% compared to the Conv error of 70.5\%. Similarly, the predicted output $\hat{f}_{mv}$ from PiLSTM shows a lower error of 39.3\% compared to the 78.9\% of Conv output.

\begin{figure}[!ht]
  \centering
  \includegraphics[width=\columnwidth]{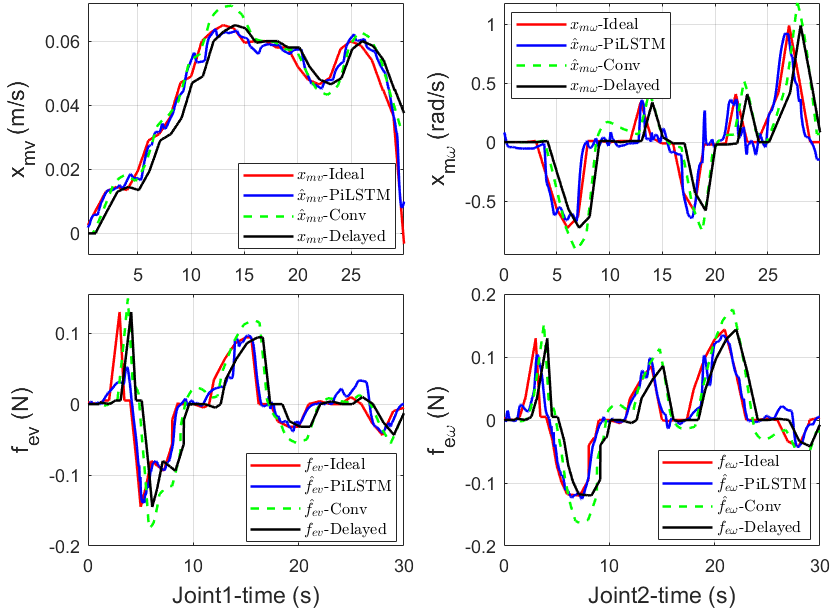}
  \caption{Operator \#2: Performance comparison of motion commands (top) and force feedbacks (bottom) among the delayed case (1.25 s), conventional predicted case, and PiLSTM predicted case in an open-loop scenario.}
  \label{perm2}
  \vspace{-10pt}
\end{figure}

\begin{figure}[!ht]
  \centering
  \includegraphics[width=\columnwidth]{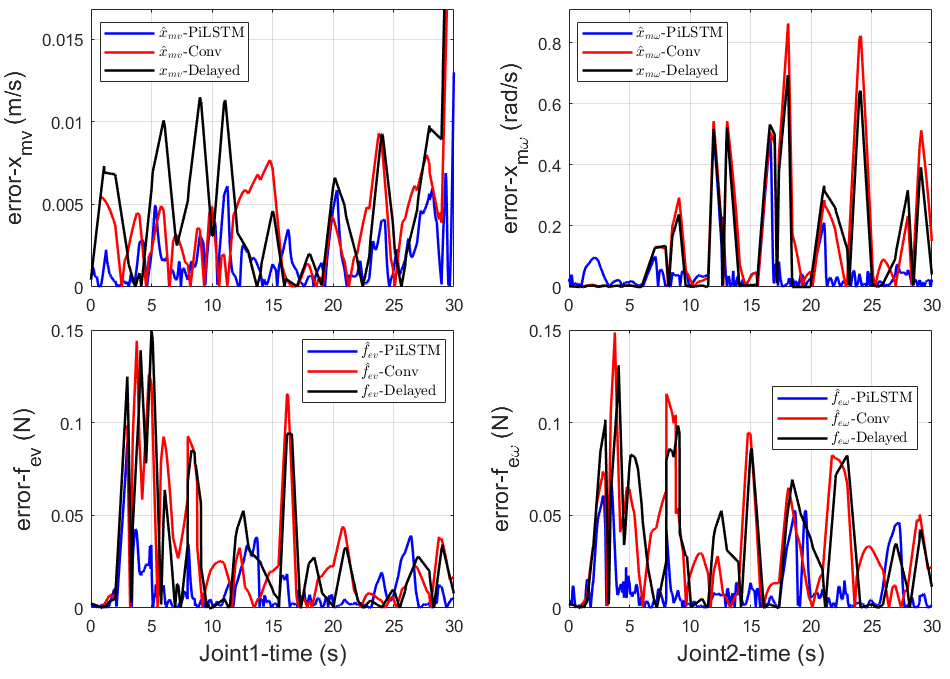}
  \caption{Operator \#2: Absolute prediction error comparison of motion commands (top) and force feedbacks (bottom) among the ideal case, delayed case (1.25 s), conventional predicted case, and PiLSTM predicted case.}
  \label{trackspic2}
  \vspace{-10pt}
\end{figure}

Similarly, performance comparisons for Operators 2, 3, 4, and 5 were conducted in the remaining experiments. The comparative analysis of the coupling variables for Operator \#2, along with the corresponding error metrics, is displayed in Fig.~\ref{perm2} and Fig.~\ref{trackspic2}, respectively. Complementing these visual representations, the normalized performance metrics for this operator and all the remaining operators, which encapsulate the delay compensation performance analysis, are summarized in Table~\ref{table3}. The best prediction accuracy for PiLSTM is 23.6\%, compared to the Conv best of 49.4\% for $\hat{x}_{mv}$, as highlighted in green, a difference of 25.8\%. Furthermore, the average performance metric for all variables (framework) for Operator \#4 is 32.3\% for PiLSTM, in contrast to Conv's 58.4\%, indicating an improvement of around 26.1\%. However, the Conv has the highest error of 116.9\% shaded in brown, indicating that the framework amplifies the negative effect of the delay at some instances instead of compensating, as highlighted in red.

In conclusion, the analysis conducted across multiple operators in open-loop scenarios consistently showed that the developed PiLSTM framework outperforms the Conv framework in delay compensation, with an average improvement of 26.1\%, thus demonstrating its enhanced capability for accurate and reliable bilateral teleoperator controls.

\renewcommand{\arraystretch}{1.1} 
\begin{table*}[ht]
\centering
\caption{Normalized performance metrics in percentage for open-loop case studies}
\begin{tabular}{|p{2.0cm}|p{1.2cm}|p{1.2cm}|p{1.2cm}|p{1.2cm}|p{1.2cm}|p{1.2cm}|p{1.2cm}|p{1.2cm}|}
\hline 
{Operators} & \multicolumn{2}{c|}{\textbf{$\Delta_n - x_{mv}(\%)$}} & \multicolumn{2}{c|}{$\Delta_n - x_{m\omega}(\%)$} & \multicolumn{2}{c|}{$\Delta_n - f_{ev}(\%)$} & \multicolumn{2}{c|}{$\Delta_n - f_{e\omega}(\%)$} \\
\cline{2-9} 
& PiLSTM & Conv & PiLSTM & Conv & PiLSTM & Conv & PiLSTM & Conv \\
\hline 
Operator \#1 & 38.3 & 70.5 & 42.5 & 68.7 & 39.2 & 78.9 & 44.5 & 89.9 \\
\hline
Operator \#2 & 28.7 & 58.0 & 37.9 & 72.1 & 33.7 & 97.8 & 35.3 & 102.7{\cellcolor{yellowCell}} \\
\hline
Operator \#3 & 54.2 & 86.3 & 50.8 & 116.9{\cellcolor{yellowCell}} & 49.7 & 88.1 & 58.1 & 98.0 \\
\hline
Operator \#4 & 23.6{\cellcolor{greenCell}} & 49.4{\cellcolor{greenCell}} & 34.5 & 53.0 & 33.4 & 69.2 & 41.7 & 78.1 \\
\hline
Operator \#5 & 52.0 & 108.3{\cellcolor{yellowCell}} & 72.5 & 82.1 & 57.4 & 79.6 & 74.2 & 91.6 \\
\hline
\end{tabular}
\label{table3}
\vspace{-5pt}
\end{table*}

\subsection{Case Study 2}
This study evaluates the teleoperation performance of the developed PiLSTM predictor frameworks, along with the blended architecture of teleoperated UGVs in closed-loop scenarios~\cite{b7a}.

The performance and transparency of the ideal case for operator \#1 in the track-following task are depicted in Fig. \ref{Ideal}. The natural deviation of the UGV states from the given motion command due to induced longitudinal slippage is well-regulated, as the operator can perceive the induced slippage and apply the necessary correction through high-fidelity closed-loop integration. As a result, the command-tracking performance becomes nearly perfect, as evidenced by the velocity tracking and force feedback tracking metrics, which is consistent with the results obtained in~\cite{b6,b12}. However, after introducing a forward and backward delay of $1.0 \sim \mathcal{U}(-0.25, 0.25)$ s each in this ideal case for the same operator—Delayed case, the system's performance and transparency significantly deteriorate, as shown by the velocity and force feedback tracking metrics in Fig. \ref{Delayed}. This decline is attributed to the degradation in the fidelity of the closed-loop integration because the operator perceives delayed force feedback-induced slippage, which in turn delays his responses to the UGV. As this effect accumulates, the system becomes less stable, leading to fast fluctuations, until the operator reduces their pace and initiates the process again, as evident at around 32 s. Thus making it difficult for the operator to control the UGV to maintain good command-tracking performance. This finding is consistent with the results in~\cite{b7a}. Furthermore, the time taken to accomplish the task increased from 148~s to 206~s.

Building upon the observed performance deterioration due to the introduced delays, we subsequently integrated the developed PiLSTM predictor framework to compensate for the delays—Predicted case. The implementation of this framework marked a significant improvement in both the performance and transparency of the system for operator \#1, as evident by the velocity and force feedback tracking metrics presented in Fig.~\ref{Predicted}. The framework effectively reduced the discrepancy between the motion commands and the actual state of the UGV, as well as the environmental force and the force perceived by the operator, by accurately predicting the undelayed commands and force feedback induced slippage. The command-tracking performance approached more stable levels observed in the ideal case, indicating a near recovery to the system's optimal state and the restoration of high-fidelity closed-loop integration. This substantial improvement underscores the efficacy of our delay compensation framework in restoring system performance. The enhanced time required to complete the task is now significantly reduced from the delayed scenario to 182~s.  Refer to supplementary material for zoomed plots.

Table IV illustrates the effectiveness of the PiLSTM framework in reducing both velocity (in m/s for $x_{mv}$ and rad/s for $x_{m\omega}$) and force feedback tracking errors (in N). The Ideal case consistently shows the lowest error values across all operators, signifying perfect performance. In contrast, the Delayed case presents relatively higher error values, highlighting the compromised fidelity of the closed-loop integration. Notably, the Predicted case shows reduced tracking error in both $x_{mv}$ and $x_{m\omega}$ compared to the Delayed case, closely approaching the performance of the Ideal case. This trend of improvement is similarly observed in the metrics $f_{ev}$ and $f_{v\omega}$, and is consistent across all operators. However, for certain cases, the PiLSTM framework exhibits large errors for $f_{v\omega}$ closer to the Delayed case, indicating some challenges in accurately predicting force feedback. This is likely due to the less predictable nature of induced slippage. For instance, $f_{v\omega}$ for operators \#2 and \#3 are approximately 18.71N and 13.23N, respectively, shaded in red. Additionally, the prediction error variation for $x_m$ is considerably smaller than for $f_e$, showing the framework is more robust in predicting motion commands than force feedback. Overall, this analysis suggests the efficacy of the developed framework in compensating for large delays and subsequently enhancing the fidelity of closed-loop integration.

The bar chart depicted in Fig.~\ref{Times} compares the track completion times across the three conditions for the five operators. Notably, the Predicted case showcases a significant reduction in completion times compared to the Delayed case, highlighting the practical benefits of the framework in delayed teleoperated UGV operations. Although the Ideal case remains the best-performing scenario, the Predicted case narrows this gap, particularly for operators 1 and 4.

\begin{figure}[!ht]
\centering
\includegraphics[width=\columnwidth]{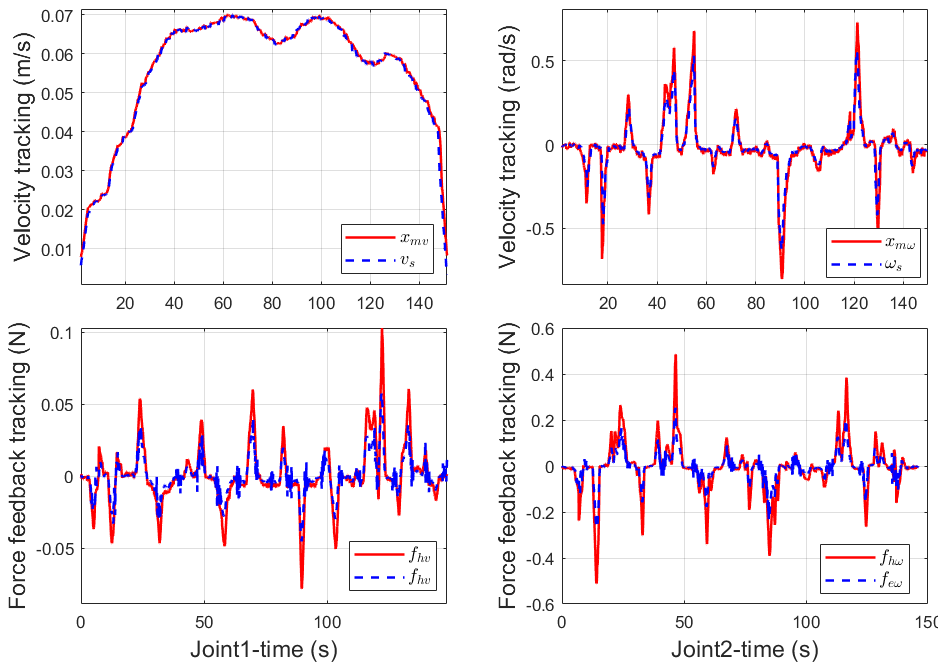}
\caption{Ideal case - Operator \#1: Performance and transparency comparison of motion commands (top) and force feedbacks (bottom), respectively. Baseline performance}
\label{Ideal}
\vspace{-5pt}
\end{figure}

\begin{figure}[!ht]
\centering
\includegraphics[width=\columnwidth]{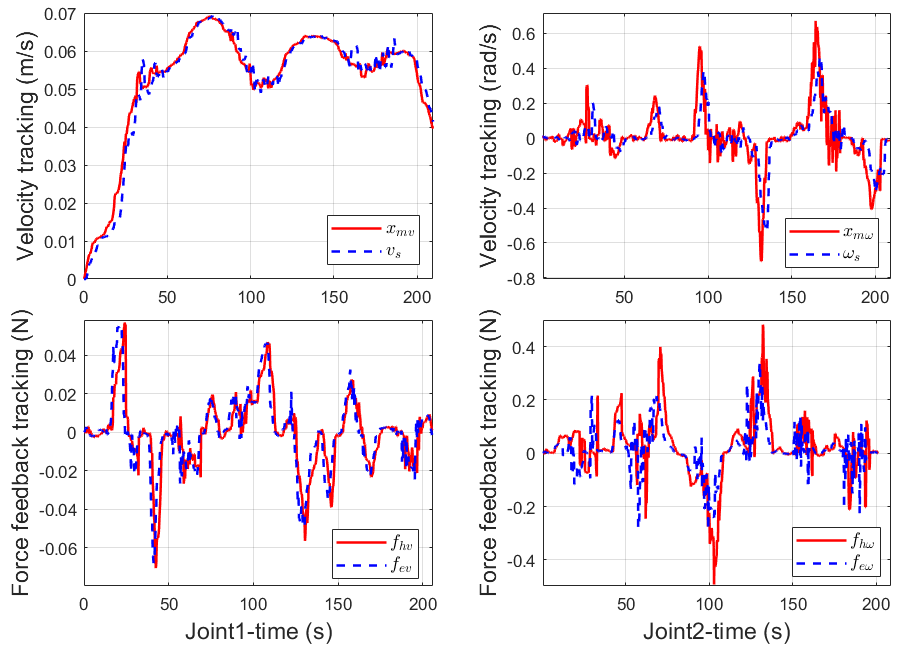}
\caption{Delayed case - Operator \#1: Performance and transparency comparison of motion commands (top) and force feedbacks (bottom), respectively. With the introduced communication delay of 1.25~s.}
\label{Delayed}
\vspace{-5pt}
\end{figure}

\begin{figure}[!ht]
\centering
\includegraphics[width=\columnwidth]{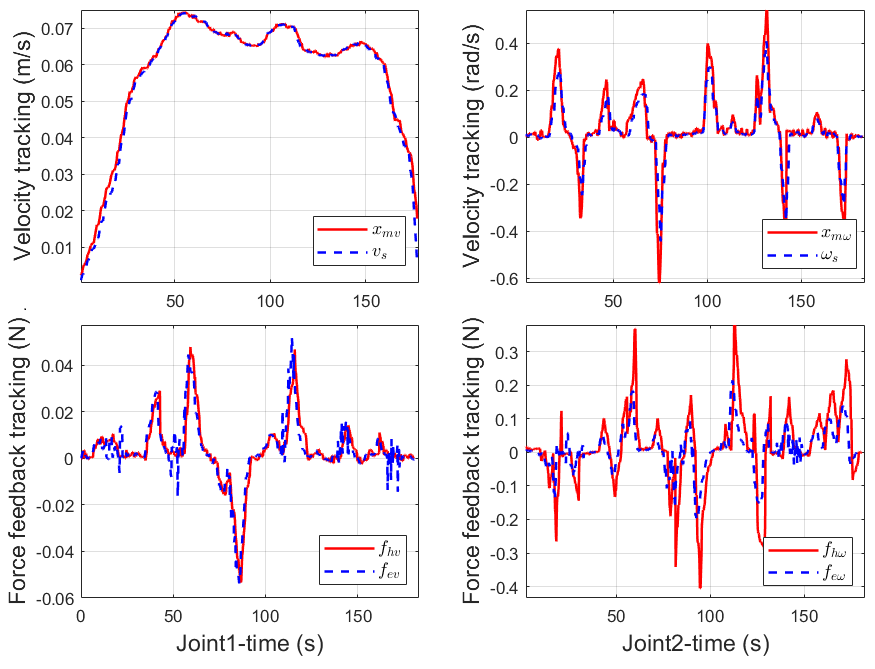}
\caption{Predicted case - Operator \#1: Performance and transparency comparison of motion commands (top) and force feedbacks (bottom), respectively. With the delays and the PiLSTM predictor framework.}
\label{Predicted}
\vspace{-10pt}
\end{figure}

\renewcommand{\arraystretch}{1.1} 
\setlength{\tabcolsep}{10pt}
\begin{table*}[ht]
\centering
\caption{Performance and Transparency Metrics for Closed-Loop Case Studies}
\begin{tabular}{|c|c|c|c|c|c|c|c|c|c|c|c|c|}
\hline Operators & \multicolumn{6}{c|}{Velocity Tracking Error - \(\Omega\)} & \multicolumn{6}{c|}{Force Feedback Tracking Error - \(\Gamma\)} \\
\cline{2-13} & \multicolumn{2}{c|}{Ideal Case} & \multicolumn{2}{c|}{Delayed Case} & \multicolumn{2}{c|}{Predicted Case} & \multicolumn{2}{c|}{Ideal Case} & \multicolumn{2}{c|}{Delayed Case} & \multicolumn{2}{c|}{Predicted Case} \\
\cline{2-13} & \(x_{mv}\) & \(x_{m\omega}\) & \(x_{mv}\) & \(x_{m\omega}\) & \(x_{mv}\) & \(x_{m\omega}\) & \(f_{ev}\) & \(f_{e\omega}\) & \(f_{ev}\) & \(f_{e\omega}\) & \(f_{ev}\) & \(f_{e\omega}\) \\
\hline Operator \#1 & 0.023 & 1.66 & 0.083 & 5.17 & 0.030 & 3.13 & 0.89 & 2.84 & 4.09 & 15.41 & 1.94 & 8.44 \\
\hline Operator \#2 & 0.020 & 1.89 & 0.077 & 6.52 & 0.054 & 4.91 & 0.91 & 3.11 & 4.17 & 16.21 & 1.08 & 13.23 \cellcolor{yellowCell} \\
\hline Operator \#3 & 0.031 & 2.01 & 0.11 & 7.61 & 0.074 & 3.96 & 1.12 & 3.17 & 5.53 & 20.72 & 2.79 & 9.65 \\
\hline Operator \#4 & 0.019 & 1.05 & 0.072 & 5.09 & 0.039 & 4.42 & 1.09 & 2.56 & 3.89 & 14.56 & 1.26 & 9.31 \\
\hline Operator \#5 & 0.030 & 1.96 & 0.12 & 7.52 & 0.081 & 6.17 & 1.14 & 4.09 & 5.99 & 21.62 & 2.84 & 18.71 \cellcolor{yellowCell} \\
\hline
\end{tabular}
\vspace{-5pt}
\end{table*}

\begin{figure}[!ht]
\centering
\includegraphics[width=\columnwidth]{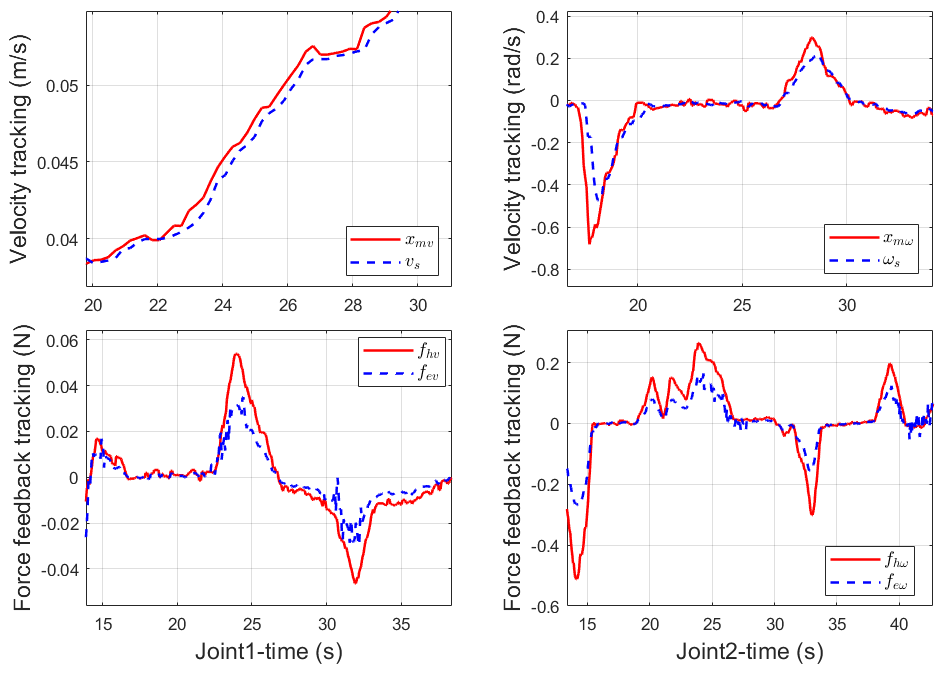}
\caption{Ideal case zoomed plot - Operator \#1: Performance and transparency comparison of motion commands (top) and force feedbacks (bottom), respectively. With perfect velocity and force feedback tracking performance, the operator was able to well perceive induced slippage through force feedback and promptly make the necessary corrections to achieve excellent command-tracking performance. Thus making the system attain high-fidelity closed-loop integration. }
\label{Ideal}
\vspace{-15pt}
\end{figure}

\begin{figure}[!ht]
\centering
\includegraphics[width=\columnwidth]{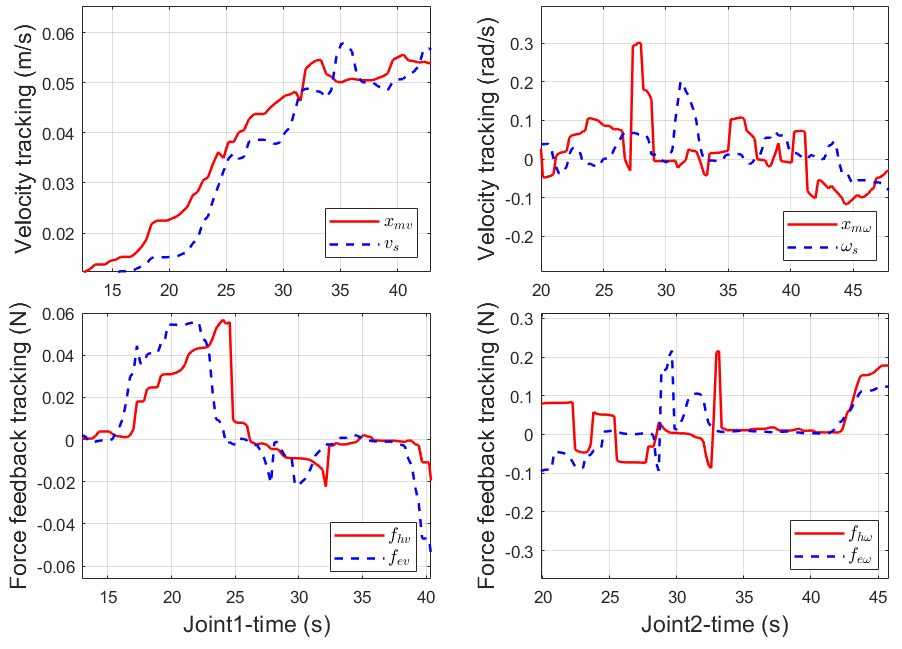}
\caption{Delayed case zoomed plot - Operator \#1: Performance and transparency comparison of motion commands (top) and force feedbacks (bottom), respectively. After introducing a communication delay of 1.25~s in both forward and backward channels, both the velocity and force feedback tracking become relatively poor, this is because the operator perceived delayed force feedback, which in turn delayed his responses. As this effect accumulates, the system becomes less stable, leading to fast fluctuations, until the operator reduces their pace and initiates the process again. Thus, poor fidelity closed-loop integration.}
\label{Delayed}
\vspace{-15pt}
\end{figure}

\vspace{-15pt}
\begin{figure}[!ht]
\centering
\includegraphics[width=\columnwidth]{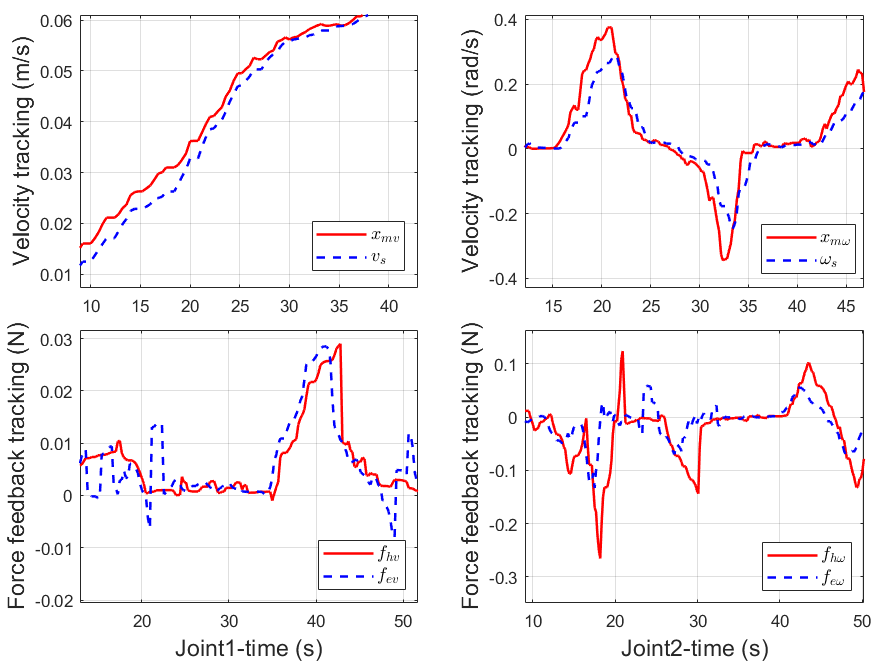}
\caption{Predicted case zoomed plot - Operator \#1: Performance and transparency comparison of motion commands (top) and force feedbacks (bottom), respectively, with the PiLSTM predictor framework. Integrating the developed framework leads to relatively better velocity and force feedback tracking, this is because the operator perceived predicted force feedback, which helped him to respond quickly.  Thus, restoring the fidelity of the closed-loop integration.}
\label{Predicted}
\vspace{-10pt}
\end{figure}

\begin{figure}[!ht]
\centering
\includegraphics[width=3.3in]{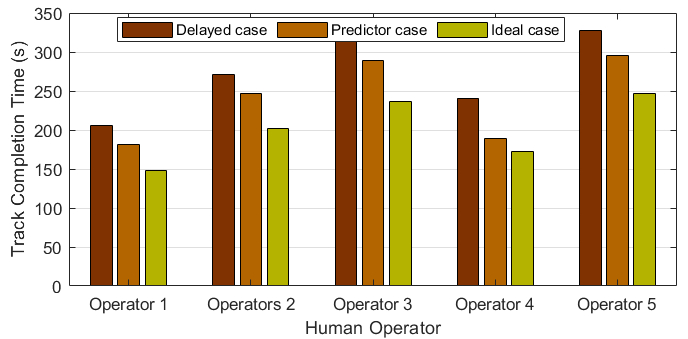}
\caption{Track completion time in delayed, predictor, and ideal cases for all operators. Delays significantly increase the completion time for all operators, and the PiLSTM predictor framework significantly improves it compared to the delayed case for operators 1 and 4.}
\label{Times}
\vspace{-2.5pt}
\end{figure}

\section{Conclusion}
In this work, a PiLSTM delay compensation framework is presented for low-speed teleoperated UGVs subjected to terrain-induced slippage and applied in a human-in-the-loop experiment. Large network delays degrade the fidelity of the closed-loop integration, making the system less stable and resulting in poor command tracking performance. We leverage the capabilities of RNNs to learn complex dynamics from data, thereby improving the prediction performance over the conventional predictor framework. Due to the inability of purely data-driven models to extrapolate complex dynamics, we integrate physical constraints to enhance the predictor's generalizations. Our results reveal that the developed PiLSTM predictor framework outperforms the conventional framework, with an improvement of around 26.1\% in open-loop studies. We further validate the framework in closed-loop studies through human-in-the-loop experiments, which proves the framework's efficacy in restoring the fidelity of the closed-loop integration, leading to a more stable system with improved command-tracking performance. Not only this, it also succeeded in reducing task completion times underlining its potential to improve the traversability and safety of teleoperated UGV for planetary exploration despite communication delays. Future work may explore a hybrid PiLSTM and model-based approach for lower prediction error in large-delay compensation.

\end{document}